\documentstyle[12pt]{article}
\input epsf.sty
\newcommand{\be}{\begin{equation}}
\newcommand{\ee}{\end{equation}}
\newcommand{\bea}{\begin{eqnarray}}
\newcommand{\eea}{\end{eqnarray}}
\newcommand{\al}{\alpha}
\newcommand{\bt}{\beta}

\newcommand{\dl}{\delta}
\newcommand{\Dl}{\Delta}

\newcommand{\kp}{\kappa}
\newcommand{\lm}{\lambda}

\newcommand{\sg}{\sigma}

\newcommand{\ta}{\tau}

\newcommand{\om}{\omega}
\newcommand{\Om}{\Omega}

\newcommand{\rarrow}{\rightarrow}
\newcommand{\nn}{\nonumber}

\begin{document}

\title{String Propagation in Bianchi Type I models: Dynamical Anisotropy 
Damping and Consequences}
\author{A. Kuiroukidis, K. Kleidis and D. B. Papadopoulos \\
Department of Physics, \\
\small Section of Astrophysics, Astronomy and Mechanics, \\
\small Aristotle University of Thessaloniki, \\
\small 54006 Thessaloniki, GREECE}

\maketitle

\begin{abstract}

A generic ansatz is introduced which provides families of 
exact solutions to 
the equations of motion and constraints for Null 
strings in Bianchi type I cosmological models. This is achieved 
irrespective of the form of the metric. Within  classes of 
dilaton cosmologies a backreaction mapping relation is established where 
the null string leads to more or less anisotropic members of the family.
The equations of motion and constraints for the generic model 
are casted in first order form 
and integrated both analytically and numerically. 

\end{abstract}
\section*{I. Introduction} 
In recent years 
string theory  has emerged as the 
most promising candidate for  the consistent quantization of
gravity and a unified description
of all the fundamental interactions. A consistent quantum theory 
of gravity is necessary part of a unified theory of all interactions, 
because pure gravity (a model containing only gravitons) cannot 
be a physical and realistic theory. This is the strongest motivation for the 
study of strings in curved spacetimes [1-3].  

In this context strings moving in curved spacetimes have gained
considerable attention, since they could provide clues to a 
proper generalization of the theory [2-3].

Among other notable features, string theory has become a theory
that gives interesting answers towards other fields, such as
the physics of Black Holes, Cosmology, Galaxy formation etc [2-4].
 
In the cosmological context inflationary models have become
natural outcomes of the theory. Yet a treatment of anisotropy 
which is a basic feature of Cosmological spacetimes has not be 
considered in the framework of string theory.

Anisotropic Cosmological spacetimes are candidates for 
a description of the early Universe, because it is known that inhomogeneous 
and anisotropic features may have prevailed in the primary stages of the 
evolution of the Universe. 

Since present day observations suggest that the Universe is highly 
homogeneous and isotropic, it would be interesting to examine
whether string theory could account for a possible altering of anisotropic 
features that may have prevailed in the primordial stages of the Universe.

Generic Bianchi I type cosmological solutions can occur in the context 
of more general dynamical theories [5]. As it is known dilaton fields 
appear naturally at the low energy limit of string theory, coupled with 
Einstein-Maxwell fields [6-8]. Also, dilaton fields appear 
as a result of dimensional reduction, of the Kaluza-Klein type Lagrangian 
[9,10]. The presence of a homogeneous primordial magnetic fields 
implies that the cosmological model is necessary anisotropic [5]. 
Much work has been done on magnetic Bianchi cosmologies in order to 
establish their main properties [11-14]. The introduction of the dilaton 
fiels renders the system in a completely integrable form [5]. 
It would be interesting to consider 
the general effects of the introduction of null strings in the above context, 
particularly in view of the possible interaction of null strings with the 
dilaton and/or EM fields and the fact that dilaton fields appear in the 
low energy-limit of string theory. 
The purpose of this paper is to study and 
examine these classes of models. 

The null string is a collection of massless point particles moving 
along the geodesics of curved spacetime [15]. Schild [16] was the first to 
consider null or tensionless strings (see also [17]). Null strings can be 
considered as the zeroth order term, 
of the perturbation expansion with respect to the string tension [18-20]. 
The evolution of null strings in different curved spacetime manifolds 
has been conidered extensively [21-23]. 

This paper is organized as follows: 
In section II we present the dynamical equations of bosonic and null 
string theory in general Bianchi type I models and 
we specify the classes of Dilaton-Bianchi cosmological 
spacetimes under consideration. \\
The generic procedure for obtaining classes of 
null-string solutions is explained in section III and   
applied to the 
specified cosmological solutions. \\
In section IV the energy-momentum tensor of string configurations is 
studied and the effects of string tension on the anisotropic features of 
these cosmological spacetimes is examined through the backreaction process. \\
The thermodynamical properties of the 
general class of solutions, through the ansatz introduced, are 
studied in section V. \\
In section VI, without further specification, the equations of motion and 
constraints for Bianchi type I models in their first order form, 
are integrated both analytically and numerically. 
\section*{II. Strings in Bianchi type I Models}\
The action for a {\it bosonic string} in a curved-spacetime background is given 
(for a D-dimensional spacetime) by [1] 
\bea
S \: = \: -\frac{T}{2} \ \int d \tau d \sigma \sqrt{-h}h^{\alpha \beta }(\tau ,\sigma )
\partial _{\alpha}X^{M} \partial _{\beta}X^{N}G_{MN}(X)
\eea
where $ \left( M,N = 0,1,...,(D-1)\right) $ are spacetime indices, $ (\alpha ,\beta = 0,1) $
are worldsheet indices and $ T = (2\pi \alpha ^{\prime })^{-1} $ is the 
string tension.

Variation of the action with respect to the "fields" which are the string 
coordinates $ X^{M}(\tau ,\sigma )$, gives the following 
equations of motion and the
constraints in the {\it conformal gauge} $\; (h_{\alpha \beta }(\tau ,\sigma ) =
exp[\phi (\tau ,\sigma )]\eta _{\alpha \beta })\; $
where $ \phi (\tau ,\sigma ) $ is an arbitrary function and 
$\; \eta _{\alpha \beta }=diag(-1,+1)\; $ [2]
\bea
\ddot{X}^{M}-(X^{M})^{\prime \prime}+\Gamma ^{M}_{AB}[\dot{X}^{A}\dot{X}^{B}
-X^{\prime A}X^{\prime B}] = 0 \\
G_{AB}(X)[\dot{X}^{A}\dot{X}^{B}+X^{\prime A}X^{\prime B}] = 0 \\
G_{AB}(X)\dot{X}^{A}X^{\prime B} = 0 
\eea
where the dot stands for $ \partial _{\tau } $ and the prime stands for
$ \partial _{\sigma } $. 

The equations of motion and constraints for the {\it null-strings} are given by 
\bea
\ddot{X}^{M}+\Gamma ^{M}_{AB}\dot{X}^{A}\dot{X}^{B}= 0 \\
G_{AB}(X)\dot{X}^{A}\dot{X}^{B}= 0 \\
G_{AB}(X)\dot{X}^{A}X^{\prime B} = 0 
\eea
The action (1) is invariant under general reparametrizations 
$\; \tilde{\sigma }^{\al }=\tilde{\sigma }^{\al }(\sigma ^{\bt })\; $ 
whereas the null string equations of motion and constraints only 
under $\; \tau _{1}=f(\tau ,\sigma ),\; \sigma _{1}=g(\sigma )\; $[15].

We consider the four-dimensional Bianchi type I 
model as the background spacetime, where the string
coordinates are denoted by $t=t(\tau ,\sigma )\; $,
$X=X(\tau ,\sigma )\; $,
$Y=Y(\tau ,\sigma )\; $,
$Z=Z(\tau ,\sigma )\; $. 

The metric tensor is given by
\be
ds^{2}=- dt^{2}+a^{2}(t)dx^{2}+b^{2}(t)dy^{2}+c^{2}(t)dz^{2} 
\ee
where $\; \al ,b,c \; $, are differentiable functions of the cosmic time. 
The Bianchi I spacetime represents an expanding universe since the volume
element of any spacelike hypersurface of simultaneity 
is constantly increasing [24,26]
\be 
\sqrt{-G}=\sqrt{-^{(3)}G}=a(t)b(t)c(t)
\ee 
The equations of motion for ${X^{\mu }=X^{\mu }(\tau ,\sigma )},
\; \; \; (\mu =0,1,2,3)$ are given 
\bea 
\ddot{t}-t^{''}&=&a\dot{a}[-(\dot{X})^{2}+(X^{'})^{2}]
+b\dot{b}[-(\dot{Y})^{2}+(Y^{'})^{2}]+\nn \\
&+&c\dot{c}[-(\dot{Z})^{2}+(Z^{'})^{2}]\\
& &\partial _{\tau }[a^{2}(t)\dot{X}]=
\partial _{\sigma }[a^{2}(t){X}^{'}]\\
& &\partial _{\tau }[b^{2}(t)\dot{Y}]=
\partial _{\sigma }[b^{2}(t){Y}^{'}]\\
& &\partial _{\tau }[c^{2}(t)\dot{Z}]=
\partial _{\sigma }[c^{2}(t){Z}^{'}]
\eea
The constraints become 
\bea
-\left[(\dot{t})^{2}+(t^{'})^{2}\right]+ 
a^{2}(t)\left[(\dot{X})^{2}+(X^{'})^{2}\right]&+&\nn \\
+b^{2}(t)\left[(\dot{Y})^{2}+(Y^{'})^{2}\right]
+c^{2}(t)\left[(\dot{Z})^{2}+(Z^{'})^{2}\right]&=&0\\ 
-\dot{t}t^{'}+
a^{2}(t)\dot{X}X^{'}
+b^{2}(t)\dot{Y}Y^{'}
+c^{2}(t)\dot{Z}Z^{'}&=&0
\eea
We will be concerned with the case 
$(t)\geq 0$ and normalize our solutions so that $(\tau \geq 0)$.

The {\it open string} boundary conditions demand
\bea
(X^{M})^{'}(\tau ,\sigma =0)=(X^{M})^{'}(\tau ,\sigma =\pi )=0 
\eea
while {\it closed strings} 
\bea
X^{M}(\tau ,\sigma =0)=X^{M}(\tau ,\sigma =2\pi )
\eea
Introducing {\it light-cone} coordinates, 
$\; \chi ^{\pm }\equiv (\tau \pm \sigma )\; $, 
the equations of motion become  
\bea
\partial _{+}\partial _{-}t&+&
\al \dot{\al }(\partial _{+}X)
(\partial _{-}X)+
b\dot{b}(\partial _{+}Y)
(\partial _{-}Y)+\nn \\
&+&c\dot{c}(\partial _{+}Z)
(\partial _{-}Z)=0\\
\al \; \partial _{+}\partial _{-}X&+&
\dot{\al }[\partial _{+}X\partial _{-}t+\partial _{+}t\partial _{-}X]=0\\
b\; \partial _{+}\partial _{-}Y&+&
\dot{b}[\partial _{+}Y\partial _{-}t+\partial _{+}t\partial _{-}Y]=0\\
c\; \partial _{+}\partial _{-}Z&+&
\dot{c}[\partial _{+}Z\partial _{-}t+\partial _{+}t\partial _{-}Z]=0 
\eea
and the constraints read 
\bea
-\left[(\partial _{+}t)^{2}\pm (\partial _{-}t)^{2}\right]+ 
a^{2}(t)\left[(\partial _{+}X)^{2}\pm (\partial _{-}X)^{2}\right]&+&\nn \\
+b^{2}(t)\left[(\partial _{+}Y)^{2}\pm (\partial _{-}Y)^{2}\right]
+c^{2}(t)\left[(\partial _{+}Z)^{2}\pm (\partial _{-}Z)^{2}\right]&=&0
\eea 
Recently, cosmological solutions of the Bianchi type I, possessing {\it dilaton} and 
{\it electromagnetic} fields have been considered [5].
The solutions consist of two families of particular solutions 
for the scale factors and the dilaton field. The action is given by 
( note however several misprints in [5] ) 
\be 
S=\int \sqrt{-g}(R+\lm g^{\mu \nu }\om _{,\mu }\om _{,\nu }
+\frac{1}{2}exp(-2\om )F_{\al \bt }F^{\al \bt })d^{4}x 
\ee 
and the equations of motion read 
\bea 
\Box \om &=&\frac{e^{-2\om }}{2\lm }F_{\mu \nu }F^{\mu \nu }\\
(e^{-2\om }F^{\mu \nu })_{;\; \; \nu }&=&0\\
G_{\mu \nu }&=&T_{\mu \nu }^{(MATT)}\\
\Box &\equiv &g^{\mu \nu }\nabla _{\mu }\nabla _{\nu } \\
T_{\mu \nu }^{(MATT)}&\equiv &
-\lm (\om _{,\mu }\om _{,\nu }-\frac{1}{2}g_{\mu \nu }\om _{,\al }\om ^{,\al })-
\nn \\
&-&exp(-2\om )(F_{\mu \al }F^{\al \nu }-\frac{1}{4}g_{\mu \nu }
F_{\al \bt }F^{\al \bt })
\eea
with $\; \om \; $ the dilaton field. 
The first family of particular solutions is given by  
\bea 
\al (\tau )&=&A_{0}e^{x_{0}\tau },\; \; \; 
b(\tau )=B_{0}e^{y_{0}\tau }\nn \\
c(\tau )&=&C_{0}e^{z_{0}\tau },\; \; \; 
\omega (\tau )=W_{1}\tau +W_{0} 
\eea 
where the constants satisfy 
$\; \lm W_{1}+2(x_{0}y_{0}+y_{0}z_{0}+z_{0}x_{0})=0\; $ with $\; \lm \; $ 
the dilaton coupling constant and   
$\; \tau \; $ defined with respect to the cosmic time t as 
$\; dt=\al (\tau )b(\tau )c(\tau )d\tau \; $. \
The second family is given by 
\bea 
\al (\tau )&=&A_{0}\tau ^{-p},\; \; \; 
b(\tau )=B_{0}\tau ^{p}\nn \\
c(\tau )&=&C_{0}\tau ^{p},\; \; \; 
\omega (\tau )=W_{0}ln \tau 
\eea 
with $\; p=\lm /(\lm-2), W_{0}=2/(\lm -2)\; $. 
We want the kinetic term of the dilaton field to be positive, 
i.e. $\; \lm <0\; $ which requires $\; 0<p<1\; $. 

There exist also three families of general solutions. 
The first one which we shall deal with, 
is given in terms of the initial expansion rates 
along the three directions $\; (\om _{1},\om _{2},\om _{3})\; $. 
Here $\; \Om _{0}\equiv \om _{1}\om _{2}+\om _{2}\om _{3}+\om _{1}\om _{3}\; $,
$\; r^{2}\equiv A^{2}(1+A^{2})\; \; $, $\; \lm =-2A^{2}\; $ 
and $\; s\equiv A^{2}\om _{1}\; $.

\bea 
\om (\tau )&=&-\frac{s}{r^{2}}\tau +\om _{0}
-\frac{A^{2}}{r^{2}}ln\left[sinh\left(\frac{\bt r}{A^{2}}\tau \right)\right]\\
\al (\tau )&=&A_{0}e^{-(sA^{2}/r^{2})\tau }
\left[sinh\left(\frac{\bt r}{A^{2}}\tau \right)\right]^{-A^{4}/r^{2}}
e^{\om _{1}\tau }\\ 
b(\tau )&=&B_{0}e^{(sA^{2}/r^{2})\tau }
\left[sinh\left(\frac{\bt r}{A^{2}}\tau \right)\right]^{A^{4}/r^{2}}
e^{\om _{2}\tau }\\ 
c(\tau )&=&C_{0}e^{(sA^{2}/r^{2})\tau }
\left[sinh\left(\frac{\bt r}{A^{2}}\tau \right)\right]^{A^{4}/r^{2}}
e^{\om _{3}\tau }\\ 
\bt ^{2}&\equiv &\Om _{0}+(s/r)^{2} 
\eea 
\section*{III. The Generic Ansatz and Exact solutions}\ 
We denote the target space coordinates of the string by 
$\; t=t(\tau ,\sigma ), X=X(\tau ,\sigma ), 
Y=Y(\tau ,\sigma ), Z=Z(\tau ,\sigma )\; $, 
and introduce the {\sl ansatz}, 
$\; t=t[ \phi ], X=X[ \phi ], 
Y=Y[ \phi ], 
Z=Z[ \phi ]\; $, for the string coordinates, 
with $\; \phi =\phi (\tau ,\sigma )\; $ 
a function to be determined. Hence we postulate an explicit 
functional-form dependence for the string coordinates. 

The metric tensor functions of Eq (8) assume the form 
\bea 
\alpha ^{2}(t)=\alpha ^{2}[t(\phi )]=\tilde{\alpha }^{2}(\phi ) \\
b^{2}(t)=b^{2}[t(\phi )]=\tilde{b}^{2}(\phi ) \\
c^{2}(t)=c^{2}[t(\phi )]=\tilde{c}^{2}(\phi ) 
\eea 
Now we can define the functions 
$\; \xi =\xi [\phi (\tau ,\sigma )],\; \eta =\eta [\phi (\tau ,\sigma )],\; 
\zeta =\zeta [\phi (\tau ,\sigma )]\; $, 
through the differential relations 
\bea 
d\xi &\equiv &\tilde{a}(\phi )dX=\tilde{a}(\phi )X_{,\phi }d\phi \\
d\eta &\equiv &\tilde{b}(\phi )dY=\tilde{b}(\phi )Y_{,\phi }d\phi \\
d\zeta &\equiv &\tilde{c}(\phi )dZ=\tilde{c}(\phi )Z_{,\phi }d\phi  
\eea 
The comma's with respect to $\; \phi \; $ denote total differentiation 
of the indicated functions, these being  
considered as single-variable functions of it. 
The constraints (22), become 
\be 
\xi _{,\phi }^{2}+\eta _{,\phi }^{2}+\zeta _{,\phi }^{2}-t_{,\phi }^{2}=0 
\ee
\\
Eqs (19)-(21) become  
\bea  
\xi _{,\phi }(\partial _{+}\partial _{-}\phi )&+&
\xi _{,\phi \phi }(\partial _{+}\phi )(\partial _{-}\phi )+\nn \\
&+&\left[\frac{\al _{,t}}{\al }\right]t_{,\phi }\xi _{,\phi }
(\partial _{+}\phi )(\partial _{-}\phi )=0\\
\eta _{,\phi }(\partial _{+}\partial _{-}\phi )&+&
\eta _{,\phi \phi }(\partial _{+}\phi )(\partial _{-}\phi )+\nn \\
&+&\left[\frac{b_{,t}}{b}\right]t_{,\phi }\eta _{,\phi }
(\partial _{+}\phi )(\partial _{-}\phi )=0\\
\zeta _{,\phi }(\partial _{+}\partial _{-}\phi )&+&
\zeta _{,\phi \phi }(\partial _{+}\phi )(\partial _{-}\phi )+\nn \\
&+&\left[\frac{c_{,t}}{c}\right]t_{,\phi }\zeta _{,\phi }
(\partial _{+}\phi )(\partial _{-}\phi )=0 
\eea 
\\
Eq (18) yields  
\bea 
t_{,\phi }(\partial _{+}\partial _{-}\phi )+
t_{,\phi \phi }(\partial _{+}\phi )(\partial _{-}\phi )
+ 
\left[\frac{\al _{,t}}{\al }\right]
(\xi _{,\phi })^{2}
(\partial _{+}\phi )(\partial _{-}\phi )+\nn \\
+
\left[\frac{b_{,t}}{b}\right]
(\eta _{,\phi })^{2}
(\partial _{+}\phi )(\partial _{-}\phi )
+
\left[\frac{c_{,t}}{c}\right]
(\zeta _{,\phi })^{2}
(\partial _{+}\phi )(\partial _{-}\phi )=0  
\eea
We consider here three classes of solutions, corresponding to the choices 
\bea  
  {\bf I.}&\; &\; (\partial _{+}\partial _{-}\phi )=0\; \\
 {\bf II.}&\; &\;(\partial _{+}\partial _{-}\phi )=\kappa (\partial _{+}\phi )
(\partial _{-}\phi )\; \\
{\bf III.}&\; &\; (\partial _{+}\partial _{-}\phi )=\mu 
\frac{t_{,\phi }}{t}(\partial _{+}\phi )(\partial _{-}\phi )
\eea
for the function $\; \phi =\phi (\chi ^{\pm })\; $. Here 
$\; \kappa ,\mu \neq 0\; $ are non-zero constants. 
The general solution of the first equation (wave equation) is given by 
\be 
\phi =\phi (\chi ^{\pm })=H(\chi ^{+})+G(\chi ^{-}) 
\ee 
while for Eq(48), 
\be 
\phi =\phi (\chi ^{\pm })=\frac{-1}{\kappa }
ln\left[H(\chi ^{+})+G(\chi ^{-})\right] 
\ee 
where H, G are arbitrary integration functions of the indicated arguments. 

{\bf I.} The equations of motion (33)-(36) and the constraint (32)
become a set of four 
second-order ordinary differential equations and a first order 
differential constraint, namely 
\bea 
t_{,\phi \phi } 
+ 
\left[\frac{\al _{,t}}{\al }\right](\xi _{,\phi })^{2}
+
\left[\frac{b_{,t}}{b}\right](\eta _{,\phi })^{2}
+
\left[\frac{c_{,t}}{c}\right](\zeta _{,\phi })^{2}=0
\eea
\bea  
\xi _{,\phi \phi }
+\left[\frac{\al _{,t}}{\al }\right]
t_{,\phi }\xi _{,\phi }=0\\
\eta _{,\phi \phi }
\left[\frac{b_{,t}}{b}\right]
t_{,\phi }\eta _{,\phi }
=0\\
\zeta _{,\phi \phi }
\left[\frac{c_{,t}}{c}\right]
t_{,\phi }\zeta _{,\phi }
=0 
\eea 
and 
\be 
\xi _{,\phi }^{2}+\eta _{,\phi }^{2}+\zeta _{,\phi }^{2}-t_{,\phi }^{2}=0 
\ee
Eqs (53)-(55) integrate to 
\bea 
\xi _{,\phi }=\frac{c_{x}}{\al },\; \; 
\eta _{,\phi }=\frac{c_{y}}{b },\; \; 
\zeta _{,\phi }=\frac{c_{z}}{c } 
\eea 
where $\; c_{x},c_{y},c_{z}\; $are arbitrary constants, 
while substituting Eqs (57) into (52), and integrating 
once, results in the same equation to be satisfied 
with the case when one 
substitutes Eqs (57) into (56),  which is 
\be 
[t_{,\phi }]^{2}=\frac{c_{x}^{2}}{\al ^{2}(t)}+
\frac{c_{y}^{2}}{b^{2}(t)}+
\frac{c_{z}^{2}}{c^{2}(t)} 
\ee 
Therefore for every given set of scale factors 
$\; \al (t),b(t),c(t)\; $ one solves the ordinary differential 
equation (58) for $\; t=t(\phi )\; $. Then through Eqs (57) 
one determines 
$\; \xi =\xi (\phi ),\eta =\eta (\phi ), \zeta =\zeta (\phi )\; $, 
with $\; \phi =\phi (\chi ^{\pm })\; $ satisfying Eq (47). 
Finally from Eqs (39)-(41) one obtains the explicit 
expressions for the string coordinates. 

{\bf II.}
Following the same line of reasoning, 
we obtain 
\bea 
\xi _{,\phi }=\frac{c_{x}}{\al }e^{-\kappa \phi },\; \; 
\eta _{,\phi }=\frac{c_{y}}{b }e^{-\kappa \phi },\; \; 
\zeta _{,\phi }=\frac{c_{z}}{c }e^{-\kappa \phi } 
\eea 
where $\; c_{x},c_{y},c_{z}\; $are arbitrary constants and 
\be 
[t_{,\phi }]^{2}=\left[\frac{c_{x}^{2}}{\al ^{2}(t)}+
\frac{c_{y}^{2}}{b^{2}(t)}+
\frac{c_{z}^{2}}{c^{2}(t)}\right]e^{-2\kappa \phi } 
\ee 
Now $\; \phi =\phi (\chi ^{\pm })\; $ satisfies eq (II) or 
equivalently Eq (51).

{\bf III.}
In the same manner, 
we obtain  
\bea 
\xi _{,\phi }=\frac{c_{x}}{t^{\mu }\al },\; \;   
\eta _{,\phi }=\frac{c_{y}}{t^{\mu }b },\; \; 
\zeta _{,\phi }=\frac{c_{z}}{t^{\mu }c } 
\eea 
where $\; c_{x},c_{y},c_{z}\; $are arbitrary constants and 
\be 
[t^{\mu }t_{,\phi }]^{2}=\left[\frac{c_{x}^{2}}{\al ^{2}(t)}+
\frac{c_{y}^{2}}{b^{2}(t)}+
\frac{c_{z}^{2}}{c^{2}(t)}\right]
\ee 
Now $\; \phi =\phi (\chi ^{\pm })\; $ satisfies Eq (49) after 
$\; t=t(\phi )\; $ as the solution of Eq (62) is substituted. 
The interesting fact is that given any set of Bianchi-type Scale-factors, 
irrespective of the general dynamical theory that they come one proceeds 
directly to a string solution depending on two arbitrary integration 
functions and seven constants. 

The invariant string size is defined as the induced metric on the 
worldsheet [3,15]
\be 
ds^{2}=G_{AB}(X)dX^{M}dX^{N}
=G_{AB}(X)X^{M}_{,\phi }X^{N}_{,\phi }(-d\tau ^{2}+d\sigma ^{2})
\ee 
which vanishes due to the constraints (22). Therefore the worldsheet 
hypersurface is {\it null} and we have classes of string solutions that are 
null-string solutions.  
It is straightforward to verify that Eqs (5)-(7) are satisfied for the 
choices of the $\; \phi =\phi (\sigma ^{\al })\; $,  
if we insert the small dimensionless parameter of the null-string perturbation 
expansion $\; c^{2}=2\lambda T\; \; \; (0<c<1)\; $ [19] 
in the terms that contain the $\; (\sigma )\; $ derivatives in Eqs (57)-(59) 
and let $\; (c\rightarrow 0)\; $. For example Eq (59) will become 
$\; (\ddot{\phi }-c^{2}\phi ^{\prime \prime })
=\mu (t_{,\phi }/t)((\dot{\phi })^{2}-c^{2}(\phi ^{\prime })^{2})\; $.

However, if we let the parameter (c) to be small $\; (c\ll 1)\; $ we can 
always reconcile the fact that the solutions are null-string solutions 
(i.e. have null worldsheet manifold) with the fact that they also satisfy 
the full bosonic string equations of motion (2)-(4).  Indeed, for this case 
while the solutions satisfy the full string equations (2)-(4), when expanded 
with respect to the parameter $\; (c)\; $ 
they also satisfy in a very good approximation the null equations (5)-(7). 
This will permit us to 
use the string energy-momentum tensor derived from the action (1) and 
study the effects of the string tension on the anisotropic features of 
spacetime, at least in the first order case with respect to the perturbation
parameter $\; (c)\; $. Thus these classes of solutions are more general than 
solutions that arise from Eqs (5)-(7). 

For the second family of particular solutions, Eqs (30), 
in the case of $\; p=+(1/2) \; $and $\; p=(1/6)\; $ 
the equations are 
integrable in closed form. One obtains the following, performing the 
integration for {\bf (I)} (and in a completely similar way 
for the cases II, III, Eqs (48)-(49) ), \\
{\bf 1.} For {\bf p=+(1/2)}
\bea 
\al (t)&=&A_{0}\left[\frac{3t}{2A_{0}B_{0}C_{0}}\right]^{-1/3}
\equiv \al _{0}t^{-1/3}\\
b(t)&=&B_{0}\left[\frac{3t}{2A_{0}B_{0}C_{0}}\right]^{1/3}
\equiv b_{0}t^{1/3}\\
c(t)&=&C_{0}\left[\frac{3t}{2A_{0}B_{0}C_{0}}\right]^{1/3}
\equiv c_{0}t^{1/3}\\
\om (t)&=&\frac{2W_{0}}{3}ln\left[\frac{3t}{2A_{0}B_{0}C_{0}}\right] 
\eea 
and for the solution 
\bea 
\left(\frac{c_{x}^{2}}{\al _{0}^{2}}t^{4/3}+c_{yz}^{2}\right)^{1/2}
&=&\left(\frac{2c_{x}^{2}}{3\al _{0}^{2}}\phi +\phi _{0}\right)\\
X_{,\phi }&=&[c_{x}/\al ^{2}(t)]\; \; \;  \\ 
Y_{,\phi }&=&[c_{y}/b^{2}(t)]\; \; \;  \\ 
Z_{,\phi }&=&[c_{z}/c^{2}(t)]\; \; \;  \\ 
c_{yz}^{2}&\equiv &(c_{y}^{2}/b_{0}^{2})+(c_{z}^{2}/c_{0}^{2})\\
t&=&t(\phi ) 
\eea 
{\bf 2.} For {\bf p=+(1/6)}
\bea 
\al (t)&=&A_{0}\left[\frac{7t}{6A_{0}B_{0}C_{0}}\right]^{-1/7}\\
b(t)&=&B_{0}\left[\frac{7t}{6A_{0}B_{0}C_{0}}\right]^{1/7}\\
c(t)&=&C_{0}\left[\frac{7t}{6A_{0}B_{0}C_{0}}\right]^{1/7}\\
\om (t)&=&\frac{6W_{0}}{7}ln\left[\frac{7t}{6A_{0}B_{0}C_{0}}\right] 
\eea 
and for the solution 
\be    
\left(\frac{c_{x}^{2}}{\al _{0}^{2}}t^{4/7}+c_{yz}^{2}\right)^{1/2}
\left(\frac{c_{x}^{2}}{\al _{0}^{2}}t^{4/7}-2c_{yz}^{2}\right)
=\frac{3c_{x}^{4}}{4\al _{0}^{4}}(\phi +\phi _{0}) 
\ee 
\bea 
X_{,\phi }&=&[c_{x}/\al ^{2}(t)]\; \; \;  \\ 
Y_{,\phi }&=&[c_{y}/b^{2}(t)]\; \; \;  \\ 
Z_{,\phi }&=&[c_{z}/c^{2}(t)]\; \; \;  \\ 
t&=&t(\phi ) 
\eea 
As an explicit demonstration of the integration procedure for the case 
$\; p=(1/2)\; $ we have the complete solution 
for the choice $\; \phi _{0}=c_{yz}\; $, as follows. We insert Eq (68) 
into Eqs (69)-(71) implicitly, through Eqs (64)-(66) and obtain 
\bea 
X(\phi )&=&\left(\frac{c_{x}}{\al _{0}^{2}}\right)
\left(\frac{4c_{x}}{3}\right)^{1/2}\left[\; \; \; 
\left(\frac{3\al _{0}^{2}}{4c_{x}}\right)
\left[2\left(\frac{c_{x}}{3\al _{0}^{2}}\right)\phi +1\right]
\sqrt{\left|\frac{c_{x}}{3\al _{0}^{2}}\phi ^{2}+\phi \right|}
\; \; \; -\right. \nn \\
&-&\left(\frac{3\al _{0}^{2}}{4c_{x}}\right)
\left(\frac{3\al _{0}^{2}}{c_{x}}\right)
ln\left(\sqrt{\left|\frac{c_{x}}{3\al _{0}^{2}}\phi +1\right|}+
\left.\sqrt{\left|\frac{c_{x}}{3\al _{0}^{2}}\phi \right|}
\; \; \right)\; \; \; \right]\\
Y(\phi )&=&
\left(\frac{2c_{y}}{b_{0}^{2}}\right)
\left(\frac{4c_{x}}{3}\right)^{-1/2}
\left(\frac{3\al _{0}^{2}}{c_{x}}\right)
ln\left(\sqrt{\left|\frac{c_{x}}{3\al _{0}^{2}}\phi +1\right|}+
\sqrt{\left|\frac{c_{x}}{3\al _{0}^{2}}\phi \right|}\; \; \; \right)\\
Z(\phi )&=&
\left(\frac{2c_{z}}{c_{0}^{2}}\right)
\left(\frac{4c_{x}}{3}\right)^{-1/2}
\left(\frac{3\al _{0}^{2}}{c_{x}}\right)
ln\left(\sqrt{\left|\frac{c_{x}}{3\al _{0}^{2}}\phi +1\right|}+
\sqrt{\left|\frac{c_{x}}{3\al _{0}^{2}}\phi \right|}\; \; \right) 
\eea 
where $\; \phi \; $ satisfies the wave equation (47). 

However we will be concerned with the most general case Eq (39) 
and keep the parameter $\; p\; $ unspecified. We have 
\bea 
\al (t)&=&A_{0}\left[\frac{(1+p)t}{A_{0}B_{0}C_{0}}\right]^{-p/(1+p)}\\
b(t)&=&B_{0}\left[\frac{(1+p)t}{A_{0}B_{0}C_{0}}\right]^{p/(1+p)}\\
c(t)&=&C_{0}\left[\frac{(1+p)t}{A_{0}B_{0}C_{0}}\right]^{p/(1+p)}\\
\om (t)&=&\frac{W_{0}}{1+p}ln\left[
\frac{(1+p)t}{A_{0}B_{0}C_{0}}\right] 
\eea 
and 
\bea 
t_{,\phi }&=&t^{-(p+\mu +p\mu )/(1+p)}\left[\left(\frac{c_{x}^{2}}{\al _{0}^{2}}
\right)t^{4p/(1+p)}+c_{yz}^{2}\right]^{1/2}\\
X_{,\phi }&=&[c_{x}/t^{\mu }\al ^{2}(t)]\\
Y_{,\phi }&=&[c_{y}/t^{\mu }b(t)]\\
Z_{,\phi }&=&[c_{z}/t^{\mu }c(t)]\\
t&=&t(\phi ) 
\eea 
The cases that are integrable in closed form 
corresponding to the above two cases are precisely 
$\; \mu =(-1+2p)/(1+p),\; $ and 
$\; \mu =(-1+6p)/(1+p)\; $, but we will retain 
the general form throughout. 

For the case  of the solution (30), from Eq (28) we get 
(following the standard procedure of [5])
\bea 
8\pi T^{(MATT)}_{00}&=&-\frac{\lm }{2}\frac{W_{0}^{2}}{(1+p)^{2}t^{2}}
-\frac{F_{0}^{2}}{2B_{0}^{2}C_{0}^{2}}
\left[\frac{A_{0}B_{0}C_{0}}{(1+p)t}\right]^{\frac{2}{(1+p)}(W_{0}+2p)}\\
8\pi T^{(MATT)}_{xx}&=&\left[-\frac{\lm }{2}\frac{W_{0}^{2}}{(1+p)^{2}t^{2}}
+\frac{F_{0}^{2}}{2B_{0}^{2}C_{0}^{2}}
\left[\frac{A_{0}B_{0}C_{0}}{(1+p)t}\right]^{\frac{2}{(1+p)}(W_{0}+2p)}
\right]\al ^{2}(t)\\
8\pi T^{(MATT)}_{yy}&=&\left[-\frac{\lm }{2}\frac{W_{0}^{2}}{(1+p)^{2}t^{2}}
-\frac{F_{0}^{2}}{2B_{0}^{2}C_{0}^{2}}
\left[\frac{A_{0}B_{0}C_{0}}{(1+p)t}\right]^{\frac{2}{(1+p)}(W_{0}+2p)}
\right]b^{2}(t)\\
8\pi T^{(MATT)}_{zz}&=&\left[-\frac{\lm }{2}\frac{W_{0}^{2}}{(1+p)^{2}t^{2}}
-\frac{F_{0}^{2}}{2B_{0}^{2}C_{0}^{2}}
\left[\frac{A_{0}B_{0}C_{0}}{(1+p)t}\right]^{\frac{2}{(1+p)}(W_{0}+2p)}
\right]c^{2}(t) 
\eea  
\section*{IV. The Energy-Momentum Tensor} 
The {\it energy-momentum} tensor for the string is obtained 
from the functional derivative of the action (1) at the 
spacetime point X. We obtain [4]
\be
\sqrt{-G}T^{AB}(X)=-\frac{T}{2}\int d\tau d\sigma 
(\dot{X}^{A}\dot{X}^{B}-X^{'A}X^{'B})\delta ^{(4)}(X-X(\tau ,\sigma )) 
\ee
where 
\be
\delta ^{(4)}(X-X(\tau ,\sigma ))=
\prod _{M=0}^{3}\delta (X^{M}-X^{M}(\tau ,\sigma ))
\ee
It is useful to integrate the energy-momentum tensor
in a three-dimensional volume that completely encompasses the 
string [3] 
\be
I^{MN}(X)=\int \sqrt{-G}T^{MN}(X)d^{3}X
\ee
For the ansatz introduced,  
$\; t=t[\phi (\tau ,\sigma )]\; , X=X[\phi ], Y=Y[\phi ], Z=Z[\phi ]\; $  
we obtain 
\be 
I^{MN}(X)=-\frac{T}{2}\int \frac{d\sigma dt(\tau ,\sigma )}{\dot{t}}
X^{M}_{,\phi }X^{N}_{,\phi }[(\dot{\phi })^{2}-(\phi ^{'})^{2}]|_{t=t[\phi ]}
\delta [t-t(\tau ,\sigma )] 
\ee
Finally using Eq (9), an {\it effective} energy-momentum tensor 
for the string is derived 
\bea 
T^{MN}_{(STRING)}(X)&=&-\frac{1}{4\pi \al ^{'}}
\frac{1}{\al (t)b(t)c(t)}
\left[\frac{X^{M}_{,\phi }X^{N}_{,\phi }}{t_{,\phi }}\right]_{t=t[\phi ]}\nn \\
& &\int d\sigma d\phi 
\frac{[(\dot{\phi })^{2}-(\phi ^{'})^{2}]}{\dot{\phi }}
\delta [\phi -\phi (\tau ,\sigma )] 
\eea 
We demonstrate the evaluation of this integral 
for the case that $\; \phi =\phi (\chi ^{\pm })\; $ 
satisfies the wave equation (47), for a closed 
string.  We have  
\bea 
\hat{I}&\equiv &\int d\sigma d\phi 
\frac{[(\dot{\phi })^{2}-(\phi ^{'})^{2}]}{(\dot{\phi })}
\delta [\phi -\phi (\tau ,\sigma )]=\nn \\
&=&\int _{0}^{2\pi }d\sigma 
\left.\frac{[(\dot{\phi })^{2}-(\phi ^{'})^{2}]}{(\dot{\phi })}
\right|_{\tau =\tau (\phi ,\sigma )} 
\eea 
where the dependance of $\; (\tau )\; $ is now in the variable 
$\; (\phi )\; $ defined by the cosmic time $\; t=t(\phi )\; $.
For closed strings satisfying the wave equation we have 
\bea  
\phi &=&\bar{\phi }+2q\al ^{'}\tau +\nn \\
&+&i\sqrt{\al ^{'}}\sum _{n\neq 0}
\frac{1}{n}[\phi _{n}e^{-in(\tau -\sigma )}+
\tilde{\phi }_{n}e^{-in(\tau +\sigma )}]
\eea 
with the reality condition $\; \phi _{n}^{*}=\phi _{-n}\; \; (n\in Z)\; $ 
and a similar relation satisfied by the tilded operators. 
We can invert approximately this relation to obtain 
$\; \tau \simeq (\phi /2q\al ^{'})\; $ and finally  
\bea 
\hat{I}=4\pi q\al ^{'}+\frac{8\pi }{q}\sum _{n>0}
Re[\phi _{n}\tilde{\phi }_{n}exp(-2in\tau )]
\eea 
In the general case, one must evaluate 
\be 
\hat{I}\equiv \int d\sigma d\phi 
\frac{[(\dot{\phi })^{2}-(\phi ^{'})^{2}]}{(\dot{\phi })}
\delta [\phi -\phi (\tau ,\sigma )]=
\int d\sigma dt(\tau ,\sigma ) 
\frac{[(\dot{\phi })^{2}-(\phi ^{'})^{2}]}{(\dot{\phi })}
\delta [t-t(\tau ,\sigma )] 
\ee 
where $\; \phi =\phi (\chi ^{\pm })\; $ satisfies Eq (39). 
This task is difficult to be achieved in its full generality.
However we can estimate the dependence on the cosmic-time$\; t\; $. 
From Eq (49), neglecting the primed factors we obtain approximately  
$\; [ln(\dot{\phi })^{2}]_{,\phi }=[lnt^{\mu }]_{,\phi }\; $ and 
since the fraction in Eq (107) is roughly proportional to 
$\; (\dot{\phi })\; $ we have $\; \hat{I}\propto t^{\mu /2}\; $. 

Therefore we obtain 
\be 
T^{MN}_{(STRING)}(t)\propto -\frac{1}{4\pi \al ^{'}}
\frac{t^{\mu /2}}{\al (t)b(t)c(t)}
\left.\left[\frac{X^{M}_{,\phi }X^{N}_{,\phi }}{t_{,\phi }}\right]
\right|_{\phi =\phi (t)} 
\ee 
With the aid of Eq (108) we 
write the functional dependence of the energy-momentum tensor 
components of the string and the "matter" content of the model that 
corresponds to Eqs (86)-(94).  
\bea 
T^{00}_{(STRING)}&\propto &\frac{
\sqrt{(c_{x}^{2}/\al _{0}^{2})t^{4p/(1+p)}+c_{yz}^{2}}t^{\mu /2}}
{t^{(p+\mu +p\mu )/(1+p)}t^{p/(1+p)}}\\ 
\nn \\
T^{0x}_{(STRING)}&\propto &\frac{t^{p/(1+p)}}{t^{\mu /2}}\\
\nn \\
T^{0y}_{(STRING)}=
T^{0z}_{(STRING)}&\propto &\frac{1}{t^{3p/(1+p)}t^{\mu /2}}\\
\nn \\
T^{xx}_{(STRING)}&\propto &\frac{ t^{(p+\mu +p\mu )/(1+p)}
t^{3p/(1+p)}}{t^{3\mu /2}
\sqrt{(c_{x}^{2}/\al _{0}^{2})t^{4p/(1+p)}+c_{yz}^{2}}}\\
\nn \\
T^{yy}_{(STRING)}=
T^{zz}_{(STRING)}&=&\nn \\
=T^{yz}_{(STRING)}
&\propto &\frac{ t^{(p+\mu +p\mu )/(1+p)}}
{t^{5p/(1+p)}t^{3\mu /2}
\sqrt{(c_{x}^{2}/\al _{0}^{2})t^{4p/(1+p)}+c_{yz}^{2}}}\\
\nn \\
T^{xz}_{(STRING)}=T^{xy}_{(STRING)}
&\propto &\frac{ t^{(p+\mu +p\mu )/(1+p)}}
{t^{p/(1+p)}t^{3\mu /2}
\sqrt{(c_{x}^{2}/\al _{0}^{2})t^{4p/(1+p)}+c_{yz}^{2}}} 
\eea  
Correspondingly, for the "matter" content of the model, they are 
given by Eqs (95)-(98) with $\; T^{xx}=\al ^{-4}(t)T_{xx} \; $, 
$\; T^{yy}=b^{-4}(t)T_{yy} \; $, $\; T^{zz}=c^{-4}(t)T_{zz} \; $, 
using the fact that $\; 0<p<1 \; $ with $\; \lm =(2p/(p-1))<0\; $ 
\bea 
8\pi T_{(MATT)}^{00}&=&\frac{pW_{0}^{2}}{(1-p)(1+p)^{2}t^{2}}
-\frac{F_{0}^{2}}{2B_{0}^{2}C_{0}^{2}}
\left[\frac{A_{0}B_{0}C_{0}}{(1+p)t}\right]^{\frac{2}{(1+p)}(W_{0}+2p)}\\
8\pi T_{(MATT)}^{xx}&=&\left[\frac{1}{A_{0}^{2}}
\frac{pW_{0}^{2}}{(1-p)(1+p)^{2}t^{2}}+\right.\nn \\
&+&\frac{F_{0}^{2}}{2B_{0}^{2}C_{0}^{2}}
\left.\left[\frac{A_{0}B_{0}C_{0}}{(1+p)t}\right]^{\frac{2}{(1+p)}(W_{0}+2p)}\right]
\left[\frac{A_{0}B_{0}C_{0}}{(1+p)t}\right]^{\frac{-2p}{(1+p)}}\\
8\pi T_{(MATT)}^{yy}&=&\left[\frac{1}{B_{0}^{2}}
\frac{pW_{0}^{2}}{(1-p)(1+p)^{2}t^{2}}\right.-\nn \\
&-&\frac{F_{0}^{2}}{2B_{0}^{2}C_{0}^{2}}
\left.\left[\frac{A_{0}B_{0}C_{0}}{(1+p)t}\right]^{\frac{2}{(1+p)}(W_{0}+2p)}\right]
\left[\frac{A_{0}B_{0}C_{0}}{(1+p)t}\right]^{\frac{2p}{(1+p)}}\\
8\pi T_{(MATT)}^{zz}&=&\left[\frac{1}{C_{0}^{2}}
\frac{pW_{0}^{2}}{(1-p)(1+p)^{2}t^{2}}\right.-\nn \\
&-&\frac{F_{0}^{2}}{2B_{0}^{2}C_{0}^{2}}
\left.\left[\frac{A_{0}B_{0}C_{0}}{(1+p)t}\right]^{\frac{2}{(1+p)}(W_{0}+2p)}\right]
\left[\frac{A_{0}B_{0}C_{0}}{(1+p)t}\right]^{\frac{2p}{(1+p)}} 
\eea  
For small values of the cosmic time $\; (t\rightarrow 0)\; $ 
or for large values, $\; (t\rightarrow +\infty )\; $,
demanding functional identification, 
between the string energy-momentum density (00) component, Eq (109), 
and the EM part of the "matter" content, Eq (115), results in 
\bea 
\mu &=&\frac{4(W_{0}+p)}{(1+p)}\; \; \; \; \; (t\rightarrow 0)\\ 
\mu &=&\frac{4(W_{0}+2p)}{(1+p)}\; \; \; \; \; (t\rightarrow +\infty )  
\eea 
The limit of $\; t\rightarrow 0\; $is taken by neglecting 
the first term in the 
square root of Eq (109), while the limit of 
$\; t\rightarrow +\infty \; $ by neglecting the second term. 
The same argument, regarding the pair of 
dilaton and EM parts of the "matter" content in Eq (115), 
yields $\; W_{0}=(1-p)\; $, 
therefore, 
\bea 
\mu &=&\frac{4}{(1+p)}\; \; \; \; \; (t\rightarrow 0)\\ 
\mu &=&4 \; \; \; \; \; \; \; \; \; \; \; (t\rightarrow +\infty )  
\eea 
For the second family of particular solutions Eqs (30), when the 
Universe is not highly anisotropic, we have $\; (p\simeq 0)\; $ 
and so the components of the string energy-momentum tensor 
Eqs (109)-(114) do not exceed the (00) component Eq (109), as 
t approaches zero. 
We can now form a sort of backreaction process, 
regarding the EM field $\; F_{0}\; $ as a small perturbing 
factor that can be ommited. The 
coefficients of the dual Universes are related by 
\bea 
\frac{p(1-p)}{(1+p)^{2}}\pm 
\left(\frac{T}{2}\right)c_{yz}=
\frac{p'(1-p')}{(1+p')^{2}} 
\eea 
In Eq (123) $\; T\equiv (1/2\pi \al ^{\prime })\; $ is the string tension. 
Using Eq (109), we assume that the presence of the null-string acts 
as an additional new source, so that the  
primed term corresponds to another member of the family, Eq (115). 
Eq (95) 
has also been used and $\; W_{0}=(1-p)\; $is substituted.  This relation 
is valid in the immediate vicinity of $\; (0\simeq p\leq 1/3)\; $. 
We obtain now the range of the cosmic-time where this coupling occurs. 
The fraction at the r.h.s. of Eq (123) is dimensionless and 
has a maximum equal to (1/8), at $\; p^{'}=(1/3)\; $. 
So the last term on the l.h.s. of Eq (123) has to be 
dimensionless and of the same order. Using Eq (90) we can therefore write 
$\; Tt^{(4+p)/(1+p)}t_{,\phi }\simeq 0.1 \; $. 

It is well known that the 
string tension can be taken to be of the order of the Planck energy scale [3]
$\; T_{Pl}=(1/2\pi \al ^{'})\simeq 10^{38}(GeV)^{2}\; $ where the Planck mass 
is given by $\; M_{Planck}=(hc/G)^{1/2}\simeq 10^{19}(GeV/c^{2})\; $ and 
correspondingly the characteristic time scale of quantum gravity is 
$\; t_{QG}\simeq 10^{-43}sec \; $while for the GUT's we have 
$\; t_{GUT}\simeq 10^{-30}sec \; $. 
The gravitational constant is $\; G\simeq 10^{-30}(GeV)^{-2}\; $. 
Since we have null strings we use in 
the action (1) the tension T to be much smaller than the one for the 
tensionful strings thus $\; (T/T_{Pl})\simeq 10^{-\kp }\; $ with 
$\; (\kp )\; $ a large integer. 

The quantity $\; t_{,\phi }\; $ has the form of inverse angular frequency, 
and for closed strings it is proportional to the modes of oscillation that 
are excited. We can therefore set equal to a integer n, (n=1,2,...).  
So we obtain 
\be 
\left(\frac{T}{T_{Pl}}\right)
\left(\frac{t}{t_{QG}}\right)^{(4+p)/(1+p)}\simeq 0.1n 
\ee 
Taking logarithms we obtain 
\be 
\left(\frac{4+p}{1+p}\right)log_{10}\left(\frac{t}{t_{QG}}\right)=
\kp -1+log_{10}n 
\ee 
This formula provides the range of cosmic-time, 
with respect to the used value of the string tension, 
where the coupling occurs. As the value of the string tension is increased 
$\; (\kp \rightarrow 0)\; $ or we have excitation of higher modes of the 
string (increasing n), the range approaches the Planckian regime.\ 
We conclude that when the string action enters with a negative sign, 
Eq (1) we get a transition to less anisotropic member of the family but 
also because of the presence of the dilaton field in this family 
of solutions it exhibits also a sort of {\it chaotic} behavior, 
in which case one can get to a more anisotropic member 
of the family 
(see Fig 1.). 
\section*{V. The General Case}\
We consider now the first family of general solutions
Eq (31)-(35). These correspond to a Universe which is singular at 
$\; t\rightarrow 0\; $, where the energy density of th magnetic field and the
scalar fields diverges and the spatial volume tends to zero [5]. 
Using Eq (12) of [5], it is not possible to obtain in closed form 
the dependance 
of the scale-factors on the cosmic-time t.
(note that it should be written as 
$\; dt=\al (\tau )b(\tau )c(\tau )d\tau )\; $. \\
However we can examine the 
cases of the {\it strong} and {\it weak} coupling limits of the Dilaton field. 
These correspond to the limits   
$\; A^{2}\rightarrow +\infty\; $ and 
$\; A^{2}\rightarrow 0\; $ respectively. We obtain  
\bea 
\lim _{A\rightarrow +\infty }
\left[sinh\left(\frac{\bt r\tau }{A^{2}}\right)\right]^{-(A^{4}/r^{2})}
&=&\frac{1}{sinh(\bt \tau )}\\
\lim _{A\rightarrow 0}
\left[sinh\left(\frac{\bt r\tau }{A^{2}}\right)\right]^{-(A^{4}/r^{2})}
&=&1 
\eea 
For the strong coupling limit we get 
\bea 
t&=&\frac{e^{\nu \tau }}{\bt ^{2}-\nu ^{2}}
[\bt cosh(\bt \tau )-\nu sinh(\bt \tau )]
+constant\\ 
\nu &\equiv &(2\om _{1}+\om _{2}+\om _{3})
\eea 
where the integration constant is chosen so that the cosmic and worldsheet
times to assume zero value simultaneously.
Now we can study the asymtotic behaviour of the solutions in two 
particular cases:\\ 
{\bf i)} 
\be 
\tau \rightarrow 0,\; \; \; t\simeq \frac{\bt }{\bt ^{2}-\nu ^{2}}
[e^{\nu \tau }-1]
\ee 
\\
Here we get 
$\; \bt =(\om _{1}\om _{2}+\om _{2}\om _{3}
+\om _{3}\om _{1})^{2}+\om _{1}^{2}\; $
\bea 
\tau &=&\frac{1}{\nu }ln\left[1+\frac{(\bt ^{2}-\nu ^{2})}{\bt }t\right]\nn \\
\al (t)&=&\frac{A_{0}}{sinh(\bt \tau )}\nn \\
b(t)&=&B_{0}e^{(\om _{1}+\om _{2})\tau }sinh(\bt \tau )\nn \\
c(t)&=&C_{0}e^{(\om _{1}+\om _{3})\tau }sinh(\bt \tau ) 
\eea 
The energy-momentum tensor for the string is given by Eq (108) together with 
Eqs (62) and (91)-(93). We now give a physical interpetation of the 
parameter $\; (\mu )\; $, in Eq (49), which is related to the thermodynamical 
interaction of the string with the geometry. 
From Eq (130), close to $\; \tau \simeq 0\; $ we have that 
$\; \tau \simeq [(\bt ^{2}-\nu ^{2})/\nu \bt ]t\; $ and the scale 
factors become 
\bea 
\al (t)&=&\frac{A_{0}\nu }{(\bt ^{2}-\nu ^{2})t}\nn \\
b(t)&=&B_{0}\frac{(\bt ^{2}-\nu ^{2})}{\nu }t\nn \\ 
c(t)&=&C_{0}\frac{(\bt ^{2}-\nu ^{2})}{\nu }t 
\eea 
From Eq (108), inside the three dimensional volume 
that completely {\it encompasses} the string, 
$\; V=\al (t)b(t)c(t)=A_{0}B_{0}C_{0}(\bt ^{2}-\nu ^{2})t/\nu \; $, 
using Eqs (101)-(103) we obtain the {\it total energy} 
$\; E^{(STR)}(t)\; $of the string and the pressures as follows 
\bea 
E^{(STR)}(t)&=&-\frac{1}{4\pi \al ^{'}}\frac{1}{t^{\mu /2}}
\left[\frac{c_{x}^{2}}{\al ^{2}(t)}+
\frac{c_{y}^{2}}{b^{2}(t)}+
\frac{c_{z}^{2}}{c^{2}(t)}\right]^{1/2}\nn \\ 
p_{x}&=&-\frac{1}{4\pi \al ^{'}}\frac{c_{x}^{2}}{t^{\mu /2}V\al ^{2}(t)}\nn \\
p_{y}&=&-\frac{1}{4\pi \al ^{'}}\frac{c_{y}^{2}}{t^{\mu /2}Vb^{2}(t)}\nn \\
p_{z}&=&-\frac{1}{4\pi \al ^{'}}\frac{c_{z}^{2}}{t^{\mu /2}Vc^{2}(t)} 
\eea 
It is well known [25] (see also [26], p.449) that if $\; F^{(STR)}(t)\; $ is the 
{\it free energy} of the string and $\; \bar{\bt }\; $the string temperature, 
we have 
\bea 
E^{(STR)}&=&F^{(STR)}+(1/\bar{\bt })S^{(STR)}\nn \\
p_{k}&=&-(\partial F^{(STR)}/\partial V)_{\bar{\bt }}
\eea 
where $\; S^{(STR)}=\bar{\bt }^{2}(\partial F^{(STR)}/\partial \bar{\bt })\; $
is the string {\it entropy}, in the three-dimensional volume. 
So if we consider {\it isothermal} motion of the string, then its 
temperature does not depend on time. Using Eq (133), 
and the fact that the volume (V) is proportional to the cosmic time t, 
we have 
\be 
\frac{dS^{(STR)}}{dt}\propto -\frac{\bar{\bt }\mu }{2}\frac{E^{(STR)}}{V} 
\ee 
so it is a measure of the entropy exchange between the string and the 
spacetime geometry. 

If we consider {\it adiabatic} motions of the three-dimensional volume 
that contains the string i.e. one allows for time dependance of the 
temperature $\; \bar{\bt }=\bar{\bt }(t)\; $ in order that there exists 
no entropy exchange then one concludes that $\; \mu \; $is a measure 
for the adjustment of the string temperature that is required for no 
entropy exchange. Using again Eq (134) we demand that there exists no 
entropy exchange $\; dS^{(STR)}/dt=0\; $ resuling in    
\be 
\frac{d\bar{\bt }(t)}{dt}\propto -\frac{\mu }{4\pi \al ^{'}}
\frac{1}{t^{\mu /2+2}}
\ee  
\section*{VI. Dynamical Isotropization and Flatness by Null Strings}\
In this Section, we integrate numerically the Einstein field equations 
which determine the evolution of a Bianchi Type I model, with matter 
content consisting solely of null-strings. A null-string may be considered 
as representing a collection of points moving independently along null 
geodesics [16]. The corresponding equations of motion and constraints are 
now given by Eqs. (5) - (7). In this case, Eq. (7) ensures that each of 
these points propagates in a direction perpendicular to the string.

In general, it is expected that the introduction of a {\em null-string 
structure} in a cosmological model will react back on the curved background, 
modifying its dynamical characteristics. In what follows, we are 
interested in determining the way that this backreaction procedure affects 
on the evolution of the Universe. Accordingly, we need to calculate the 
components of the energy-momentum tensor attributed to a spacetime region 
due to the propagation of a null-string, in an otherwise vacuum cosmological 
model, with metric in the form of Eq. (8). These quantities are subsequently 
inserted into the r.h.s. of the corresponding field equations. The resulting 
dynamical system is highly non-linear and therefore, solutions can be obtained 
only through certain numerical techniques, where the concept of {\em attractor} 
plays an important role: If some special spacetime represents an attractor for 
a wide range of initial conditions, such a spacetime is naturally realized 
asymptotically.

In a Bianchi Type I spacetime, the equations of motion for a null-string in 
the zeroth order approximation to $c^2 = 2 \lm T$, are
\be
\ddot{t} + a \: {d a \over d t} \: {\dot{x}}^2 + b \: {d b \over d t} 
\: {\dot{y}}^2 + c \: {d c \over d t} \: {\dot{z}}^2 = 0
\ee
\bea
\ddot{x} & + & 2 \: {1 \over a} \: {d a \over d t} \: \dot{t} \dot{x} = 0 
\nn \\
\ddot{y} & + & 2 \: {1 \over b} \: {d b \over d t} \: \dot{t} \dot{y} = 0 
\nn \\
\ddot{z} & + & 2 \: {1 \over c} \: {d c \over d t} \: \dot{t} \dot{z} = 0 
\eea
where a dot denotes derivative with respect to the world-sheet coordinate 
$\ta$. The corresponding constraints [Eqs. (6) and (7)] are written in the 
form
\bea
{\dot{t}}^2 & = & a^2 {\dot{x}}^2 + b^2 {\dot{y}}^2 + c^2 {\dot{z}}^2 
\nn \\
\dot{t} t^{\prime} & = & a^2 {\dot{x}} x^{\prime} + b^2 {\dot{y}} y^{\prime} 
+ c^2 {\dot{z}} z^{\prime} 
\eea
where the prime denotes differentiation with respect to $\sg$. Eqs. (138) 
are evaluated as follows
\bea
\dot{x} & = & {x_0 (\sg) \over a^2} \nn \\
\dot{y} & = & {y_0 (\sg) \over b^2} \nn \\
\dot{z} & = & {z_0 (\sg) \over c^2} 
\eea
In this case, provided that $\dot{t} \neq 0$, from Eq. (137) we deduce
\be
t^2 = t_0^2 (\sg) + {x_0^2 (\sg) \over a^2} + {y_0^2 (\sg) \over b^2} + 
{z_0^2 (\sg) \over c^2}
\ee
from which, by virtue of Eq. (139), we obtain $t_0 (\sg) = 0$. Now, with the 
aid of Eqs. (140) and (141), we may determine the components of the {\em total 
energy-momentum tensor} [Eq. (101)] attributed to a spacetime region due to 
the propagation of a null-string in it. As regards the corresponding $(00)$ 
component, we have
\be
I^{00} = - {T \over 2} \: \int d \sg \: \int d \ta \: {\dot{t}}^2 \: 
\dl ( t - t^{\prime} )
\ee
In Eq. (142), we perform the substitution $(\ta , \sg) \rarrow [ t(\ta , \sg) , 
\sg ]$ in a way such that
\be
dt \: d \sg = \dot{t} \: d \ta d \sg
\ee
thus obtaining 
\be
I^{00} = - {T \over 2} \: \int d \sg \: \int d t \: \dot{t} \: 
\dl ( t - t^{\prime} )
\ee
Now, by virtue of Eq. (141) and the identity
\be
\int d \mu (k) \: f (k^{\prime}) \: \dl ( k - k^{\prime} ) = f (k)
\ee
Eq. (144) is finally written in the form
\be
I^{00} = \mp \: {T \over 2} \: \int_0^{2 \pi} d \sg \: {1 \over a b c} \: 
\sqrt{x_0^2 \: b^2 c^2 + y_0^2 \: c^2 a^2 + z_0^2 \: a^2 b^2}
\ee
In the same fashion, we obtain
\bea
I^{01} & = & \mp \: {T \over 2} \: \int_0^{2 \pi} {x_0 (\sg) \over a^2 (t)} \: 
d \sg = I^{10} \nn \\
I^{02} & = & \mp \: {T \over 2} \: \int_0^{2 \pi} {y_0 (\sg) \over b^2 (t)} \: 
d \sg = I^{20} \nn \\
I^{03} & = & \mp \: {T \over 2} \: \int_0^{2 \pi} {z_0 (\sg) \over c^2 (t)} \: 
d \sg = I^{30}
\eea
\bea
I^{12} & = & \mp \: {T \over 2} \: \int_0^{2 \pi} x_0 y_0 \: {c \over a b} \: 
{1 \over \sqrt{x_0^2 \: b^2 c^2 + y_0^2 \: c^2 a^2 + z_0^2 \: a^2 b^2}} \: 
d \sg = I^{21} \nn \\
I^{13} & = & \mp \: {T \over 2} \: \int_0^{2 \pi} x_0 z_0 \: {b \over a c} \: 
{1 \over \sqrt{x_0^2 \: b^2 c^2 + y_0^2 \: c^2 a^2 + z_0^2 \: a^2 b^2}} \: 
d \sg = I^{31} \nn \\
I^{23} & = & \mp \: {T \over 2} \: \int_0^{2 \pi} y_0 z_0 \: {a \over b c} \: 
{1 \over \sqrt{x_0^2 \: b^2 c^2 + y_0^2 \: c^2 a^2 + z_0^2 \: a^2 b^2}} \: 
d \sg = I^{32} 
\eea
and
\bea
I^{11} & = & \mp \: {T \over 2} \: \int_0^{2 \pi} x_0^2 \: {b c \over a^3} \: 
{1 \over \sqrt{x_0^2 \: b^2 c^2 + y_0^2 \: c^2 a^2 + z_0^2 \: a^2 b^2}} \: 
d \sg \nn \\
I^{22} & = & \mp \: {T \over 2} \: \int_0^{2 \pi} y_0^2 \: {c a \over b^3} \: 
{1 \over \sqrt{x_0^2 \: b^2 c^2 + y_0^2 \: c^2 a^2 + z_0^2 \: a^2 b^2}} \: 
d \sg \nn \\
I^{33} & = & \mp \: {T \over 2} \: \int_0^{2 \pi} x_0^2 \: {a b \over c^3} \: 
{1 \over \sqrt{x_0^2 \: b^2 c^2 + y_0^2 \: c^2 a^2 + z_0^2 \: a^2 b^2}} \: 
d \sg 
\eea
Taking into account a circular (planar) string, namely
\be
x_0 (\sg) = cos \sg \; , \; \; y_0 (\sg) = sin \sg \; , \; \; z_0 = const.
\ee
which propagates along the anisotropic $z-$direction of an axisymmetric 
Bianchi Type I model, i.e.
\be
a (t) = b(t)
\ee
we obtain $t \neq t (\sg)$. Then, the only non-zero components of 
the corresponding total energy-momentum tensor are
\bea
I^{00} & = & \mp {\pi T \over a c} \: \sqrt{c^2 + z_0^2 a^2} \nn \\
I^{11} & = & \mp {\pi T \over 2 a^3} \: {c \over \sqrt{c^2 + z_0^2 a^2}} 
= I^{22} \nn \\
I^{33} & = & \mp {\pi T \over c^3} \: z_0^2 \: {a \over \sqrt{c^2 + z_0^2 a^2}} 
\eea
and
\be
I^{03} = \mp {\pi T \over c^2} \: z_0 = I^{30}
\ee
Due to the homogeneity of the spacetime under consideration, we may impose 
that $z_0 = 0$ [29]. In this case, the resulting total energy-momentum tensor 
is {\em diagonal}
\bea
I^{00} & = & \mp {\pi T \over a (t)} \nn \\
I^{11} & = & \mp {\pi T \over 2 a^3 (t)} = I^{22} \nn \\
I^{33} & = & 0 
\eea
Now, the corresponding physical quantity to be inserted into the r.h.s. of 
the field equations, may be defined as: $ \left [ T^{\mu \nu} \right ] = \left 
[ I^{\mu \nu} \: per \: unit \: of \: proper - comoving \: volume \right ]$, 
thus resulting to 
\bea
T^{00} & = & \mp {\pi T \over c a^3} \nn \\
T^{11} & = & \mp {\pi T \over 2 c a^5} = T^{22} \nn \\
T^{33} & = & 0 
\eea

In this respect, the Einstein field equations which determine the evolution 
of a Bianchi Type I model with matter in the form of null-strings, are 
written in the form
\bea
{1 \over a^2} \: ({d a \over dt})^2 + 2 \: {1 \over a} \: {d a \over d t} \: 
{1 \over c} \: {d c \over d t} & = & - 4 \pi \: ({G \over \al^{\prime}}) \: 
{1 \over a^3 c} \\ 
{1 \over a} \: {d^2 a \over dt^2} + {1 \over c} \: {d^2 c \over d t^2} + 
{1 \over a} \: {d a \over dt} \: {1 \over c} {d c \over dt} & = & 2 \pi \: 
({G \over \al^{\prime}}) \: {1 \over a^7 c} \\
2 \: {1 \over a} \: {d^2 a \over dt^2} + {1 \over a} \: {d a \over dt} \: 
{1 \over c} \: {d c \over d t} & = & 0
\eea
For the purpose of numerical analysis, in what follows we shall treat the 
dimensionless constant ${G \over \al^{\prime}}$ (where $\al^{\prime}$ is 
Regge slope) as a free parameter, to be varied at will. In fact, this 
parameter may be used to denote the regime at which gravitational phenomena 
dominate over the string effects $({G \over \al^{\prime}} > 1)$ and vice 
versa $({G \over \al^{\prime}} < 1)$.

In principle, we may integrate the system of Eqs. (156) - (158) to obtain 
the exact form of the unknown scale functions, $a(t)$ and $c(t)$, thus 
determining the evolution of the cosmological model under consideration. 
Since this is not an easy task, we may get a good estimation of their 
dynamic behaviour through numerical integration. From Eqs. (156) - (158) 
only two are truly independent. The third one corresponds to an additional 
constraint to be satisfied by the solutions of this system. As such, we 
choose Eq. (156). The remaining independent field equations (157) and (159) 
may be recast in the form of a first order system, as follows
\bea
{d H_1 \over d t} & = & - H_1^2 - {1 \over 2} \: H_1 \: H_3 \\
{d H_3 \over d t} & = & - 4 \pi \: ({G \over \al^{\prime}}) \: a^7 \: c 
- H_3^2 - {1 \over 2} \: H_1 \: H_3 
\eea
and
\be
{d a \over d t} = a \: H_1 \; \; \; , \; \; \; \; {d c \over d t} = c \: H_3
\ee
where $H_1$ and $H_3$ are the anisotropic Hubble parameters. Accordingly, 
in what follows, we integrate numerically the system of Eqs. (159) - (161). 

To avoid possible implications of small variations on the dynamical parameters 
involved, the constraint (156) is checked to be satisfied with an accuracy of 
$10^{-10}$ along numerical integration. Both Hubble parameters are measured 
in units of $10^{-4}$ $sec^{-1}$, being normalized with respect to $\sqrt{G}$. 
The initial conditions imposed on the dynamical variables of the problem 
$(H_1^0, \; H_3^0, \; a_0, \; c_0)$, are chosen so that: {\bf (a)} $a_0 = 
c_0$, i.e. at the origin, the two factor spaces (the isotropic $xy$-plane 
and the anisotropic $z$-direction) are separated, but of the same {\em linear 
dimension}. {\bf (b)} $H_1^0 > 0$, i.e. initially the isotropic space expands, 
in accordance to what we observe at the present epoch. As regards the 
anisotropic direction it may be either expanding or contracting, a thing that 
depends on the exact value of ${G \over \al^{\prime}}$, i.e. on the regime at 
which we are working at (classical or quantum).

The time coordinate is measured in dimensionless units, being normalized with 
respect to the Planck time, $t \rarrow t/t_{Pl}$ ($t_{Pl} = \sqrt {G} \sim 
10^{-43} sec$). The limits of numerical integration range from $t_0 = 10$ to 
$t_f = 10^5$. The upper limit coincides with the origin of the GUTs epoch, 
$t_{GUT} = 10^5 \: t_{Pl}$, corresponding to the end of the string era. On the 
other hand, the lower limit $(t_0 = 10 \: t_{Pl})$ is chosen as being safely 
away from the Planck epoch, since, in the absence of a self-consistent quantum 
theory of gravity, there is always a region of ambiguity around $t = 0$, of the 
order of the Planck time.

The solutions to the system of differential equations (159) - (161) may be 
represented as curves in a $H_1 - H_3$ plane. Any point located on them, 
always satisfies the constraint condition (156), as well. Thus, these curves 
actually represent {\em orbits} of the dynamical system under study. Each 
curve, corresponding to a different set of initial conditions, is bounded 
by fixed points (or infinities) and represents a different type of evolution 
for the Universe. In what follows, we focus attention on the existence and 
the evolution of attracting points in the $H_1 - H_3$ plane. The reason rests 
in the physical meaning of the attractor: No matter what the behaviour of a 
cosmological model at the origin might be, it will always end up to evolve 
as indicated by the location of the attracting point in the $H_1-H_3$ plane.

In order to study a possible {\em dissipation} of the existing anisotropy, 
we furthemore define the so called {\em anisotropy measure} [30], as 
\be
{\Dl H \over H} = {3 \over 2} \: {H_1 - H_3 \over 2 H_1 + H_3}
\ee
This dimensionless quantity is a measure of the variance in the expansion 
rates at different directions and hence it represents the degree of anisotropy. 
Accordingly, we examine its dynamical behaviour during numerical integration. 
We expect that, at the end of the backreaction process, $\Dl H / H$ must settle 
down to a constant value that approaches zero, corresponding to a {\em dynamical 
isotropization} of the anisotropic background.

Depending on the exact value of the free parameter ${G \over \al^{\prime}}$, 
we consider three different cases:

\subsection*{(i) Gravitational phenomena are comparable to string effects}

In this case, we have ${G \over \al^{\prime}} = 1$. The time evolution of the 
Hubble parameters for a particular set of initial conditions, $(H_1^0 , H_3^0) 
= (1 , -1.5)$, is presented in Fig. 2a. The cosmic-time coordinate is measured 
in units of $10^3 \: t_{Pl}$. We see that, although at the begining both Hubble 
parameters evolve anisotropically, after some time which is approximately of the 
order $\Dl t \approx 8 \times 10^3 \: t_{Pl}$, their evolution becomes identical. 
In the later stages (for $t \geq 3 \times 10^4 \: t_{Pl}$), both evolution rates 
are reduced to a static value approximately equal to zero. In this respect, a 
complete isotropization of the model is achieved (i.e. ${\Dl H \over H} = 0$) 
quite rapidly (see Fig. 2b). Notice that the evolution rate of the anisotropic 
direction $(H_3)$ never exceeds in value the corresponding parameter of the 
isotropic space $(H_1)$.

The above results indicate that, during the cosmological evolution there 
exists an attracting point in the $H_1 - H_3$ plane. The explicit location 
of this point is presented in Fig. 3. We observe that, for a wide range of 
initial conditions (including both expansion and contraction of the 
anisotropic direction), every orbit of the dynamical system under 
consideration ends up to a fixed point very close to $(0 , 0)$, i.e. the 
Minkowski spacetime. This point is stable and according to the definition of 
the attractor, the cosmological model under study is not only {\bf isotropic}, 
but also asymptotically (as $t \rarrow \infty$) {\bf flat}.

\subsection*{(ii) Gravity dominates over string effects}

In this case, we take ${G \over \al^{\prime}} = 10$. The time evolution of 
the Hubble parameters corresponding to the initial conditions of the previous 
case, is presented in Fig. 4a. The cosmic-time coordinate is also measured in 
units of $10^3 \: t_{Pl}$. We see that at the end of the backreaction process, 
an initially large degree of anisotropy is subsequently minimized, so that, 
asymptotically, the evolution rates of both factor spaces are reduced to the 
static value $(0,0)$. However, in this case, the anisotropy dissipation rate 
is quite lower and the isotropization of the model is completed only after a 
period of $\Dl t > 3 \times 10^4 \: t_{Pl}$ (see Fig. 4b). We furthermore 
observe that, although the dynamical behaviour of $H_1$ is more or less the 
same as in the previous case, the anisotropic direction undergoes a period 
$(\approx 10^2 \: t_{Pl})$ of accelerating (inflationary) expansion, during 
which the value of $H_3$ exceeds $H_1$.

In this case also, we verify the existence of an attractor in the $H_1 - H_3$ 
plane. The explicit location of this point is presented in Fig. 5. Notice that 
the dynamical behaviour of the various orbits is almost identical (there is no 
spreading of the orbits in the $H_1 - H_3$ plane), although they correspond to 
quite different sets of initial conditions. Again, the attractor coincides to 
the stable point $(0 , 0)$, i.e. the Minkowski spacetime. Therefore, the 
cosmological model under consideration is also asymptotically flat.

\subsection*{(iii) String effects dominate over gravity}

In this case, we take ${G \over \al^{\prime}} = 0.1$. The corresponding results 
are more or less similar to the previous ones, with only one major difference: 
We do not have orbits corresponding to negative initial conditions of $H_3$. 
For $H_3^0 < 0$, the constraint equation (156) is not satisfied any more. 
Accordingly, a {\em dominant} null-string structure does not permit contraction 
of the anisotropic dimension and therefore, the cosmological model under 
consideration does not admit a Kasner-type solution. Again, we verify the 
existence of an attractor in the $H_1 - H_3$ plane, which coincides to the 
stable point $(0 , 0)$ corresponding to Minkowski spacetime, while anisotropy 
dissipation occurs rather rapidly (see Figs. 6 and 7). 

In concluding, we may state that the introduction of a null-string structure 
in a Bianchi Type I spacetime, may act in favour of both isotropization and 
flatness of the anisotropic background. Moreover, in the case where the string 
effects are either comparable or dominant with respect to the gravitational 
ones, the dynamical behaviour of $H_3$ indicates that the string actually 
decelerates the evolution of the anisotropic direction (e.g. see Figs. 2a 
and 4a), thus serving as a factor for {\em counter-inflation}, as well (in 
connection see [26]).

\section*{VII. Discussion and Conclusions}\

In the first part of this paper the introduction of a generic ansatz 
has provided classes of exact, null-string solutions in a Bianchi Type I 
spacetime. This can be carried out irrespectively of the dynamical theory 
under which thecorresponding cosmological solutions may originate and in 
fact, one has to use only the explicit functional form of the scale factors. 
This procedure is applied explicitly to the case of dilaton-Bianchi Type I 
cosmological models. 

We may realize the possibility of backreaction of the strings on the curved 
background, by assuming that the dilaton and electromagnetic fields have the 
same cosmic-time dependence with the string. This specifies the values of 
the parameter $\mu$, in such a way, that allows transition to a less or more 
anisotropic member of the same family of the cosmological solutions. The 
range of values of the cosmic-time parameter in which this process is valid, 
depends on the exact value of the string tension and the number of the string 
modes that are excited, in the first order approximation. If the three 
dimensional volume that encompasses the string is considered as a whole, its 
interaction with the spacetime geometry has a rate which is proportional to 
the parameter that characterizes the general class of solutions. 

Finally, if one assumes that in the primordial stages of the evolution of 
the Universe, high spacetime curvature resulted in particle production [27], 
then one is led to examine if string-like properties of particles can also 
lead to anisotropy damping. In this respect, the  strings, being considered 
as objects carrying stress and energy, seem to provide the means for 
anisotropy damping in epochs where the spacetime could have different 
expansion rates along different directions. This is examined in the last 
part of the paper, where the cosmological field equations regarding a 
null-string distribution propagating in a Bianchi type I spacetime, are 
integrated without any further assumptions. This integration indicates that, 
for a wide range of initial conditions, the null-string structure acts in 
favour of both the isotropization and the flatness of an, initially, curved 
anisotropic background.

{\bf Acknowledgements:} One of us (A.K) would like to thank the Greek State 
Scholarships Foundation (I.K.Y), for financial support during this work. 
Financial support from the Greek Secretariat of Research and Technology 
under contract PENED-1768 is gratefully acknowledged.

\newpage

\section*{Figure Captions}

\begin{itemize}

\item {\bf Figure 1:} Plot of the function $f(p) = p(1-p)/(1+p)^{2}$

\item {\bf Figure 2a:} The time evolution of the anisotropic Hubble 
parameters $H_1$ (squa-res) and $H_3$ (solid line), for a particular 
set of initial conditions $[(H_1^0, H_3^0) = (1, -1.5)]$, in the case 
where ${G \over \al^{\prime}} = 1$. The time coordinate is measured 
in units of $10^3 \: t_{Pl}$. Notice that, the cosmological model 
becomes both isotropic (for $t \geq 8 \times 10^3 \: t_{Pl}$) and 
flat (as $t \rarrow \infty)$.

\item {\bf Figure 2b:} Plot of the anisotropy measure $({\Dl H \over H})$ 
versus time (in units of $10^3 \: t_{Pl}$), as regards the evolution of the 
Hubble parameters corresponding to the set of initial conditions $(H_1^0, 
H_3^0) = (1, -1.5)$, in the case where ${G \over \al^{\prime}} = 1$. The 
isotropization of the model is completed after $8 \times 10^3 \: t_{Pl}$.

\item {\bf Figure 3:} The orbits of the dynamical system determined by the 
cosmological field equations, for five different sets of initial conditions, 
in the $H_1 - H_3$ plane. All orbits approach asymptotically at the attracting 
point $(0 , 0)$.

\item {\bf Figure 4a:} The time-evolution of $H_1$ (squares) and $H_3$ 
(solid line), for the same set of initial conditions as in Fig. 2a, in 
the case where ${G \over \al^{\prime}} = 10$. The time coordinate is 
measured in units of $10^3 \: t_{Pl}$. Notice that in this case, the 
anisotropic direction undergoes a period of accelerating (inflationary) 
expansion before the model becomes both isotropic (for $t > 3 \times 
10^4 \: t_{Pl}$) and flat (as $t \rarrow \infty)$.

\item {\bf Figure 4b:} Similar to Figure 2b, except that ${G \over 
\al^{\prime}} = 10$. Notice that the isotropization of the model is 
quite slower in this case.
 
\item {\bf Figure 5:} Similar to Figure 3, except that ${G \over 
\al^{\prime}} = 10$. Again, all orbits end at the attracting point 
$(0 , 0)$.

\item {\bf Figure 6a:} The time evolution of $H_1$ (squares) and $H_3$ 
(solid line) for a particular set of initial conditions $[(H_1^0, H_3^0) 
= (1 , 2)]$, in the case where ${G \over \al^{\prime}} = 0.1$. The time 
coordinate is measured in units of $10^3 \: t_{Pl}$. Notice that the 
dominant null-string acts in favour of a counter-inflation of the anisotropic 
dimension, thus resulting to a rapid isotropization of the cosmological model.

\item {\bf Figure 6b:} Plot of the anisotropy measure $({\Dl H \over H})$ 
versus time (in units of $10^3 \: t_{Pl}$), as regards the evolution of the 
Hubble parameters corresponding to the set of initial conditions $(H_1^0, 
H_3^0) = (1, 2)$, in the case where ${G \over \al^{\prime}} = 0.1$. 

\item {\bf Figure 7a:} Similar to Figure 6a, except that $(H_1^0, H_3^0) = 
(1 , 0)$.

\item {\bf Figure 7b:} Similar to Figure 6b, except that $(H_1^0, H_3^0) = 
(1 , 0)$.

\end{itemize}
\newpage 
\begin{figure} 
\centerline{\mbox {\epsfxsize=15.cm \epsfysize=15.cm \epsfbox{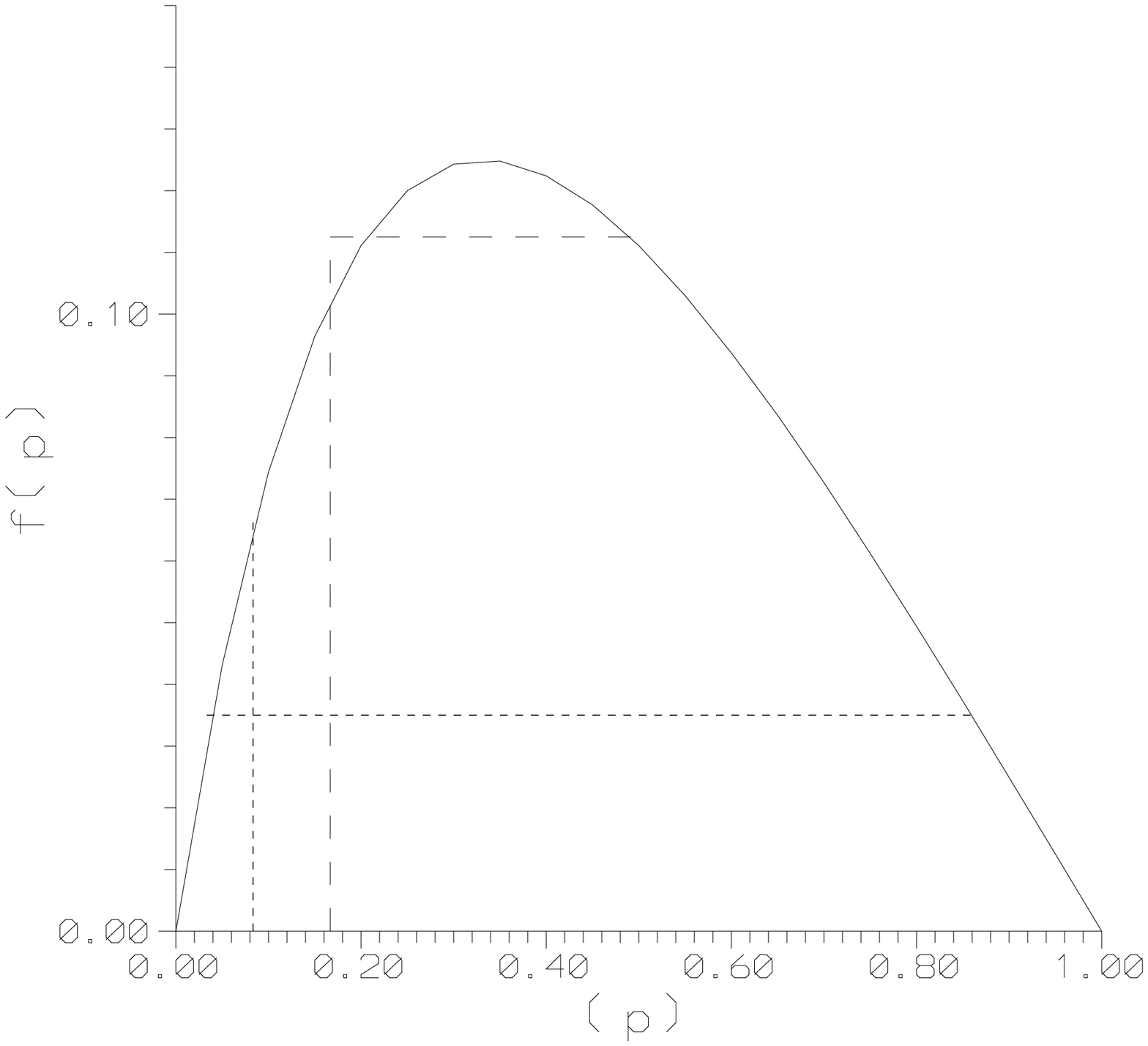}}}
\end{figure} 
\begin{figure} 
\centerline{\mbox {\epsfxsize=15.cm \epsfysize=15.cm \epsfbox{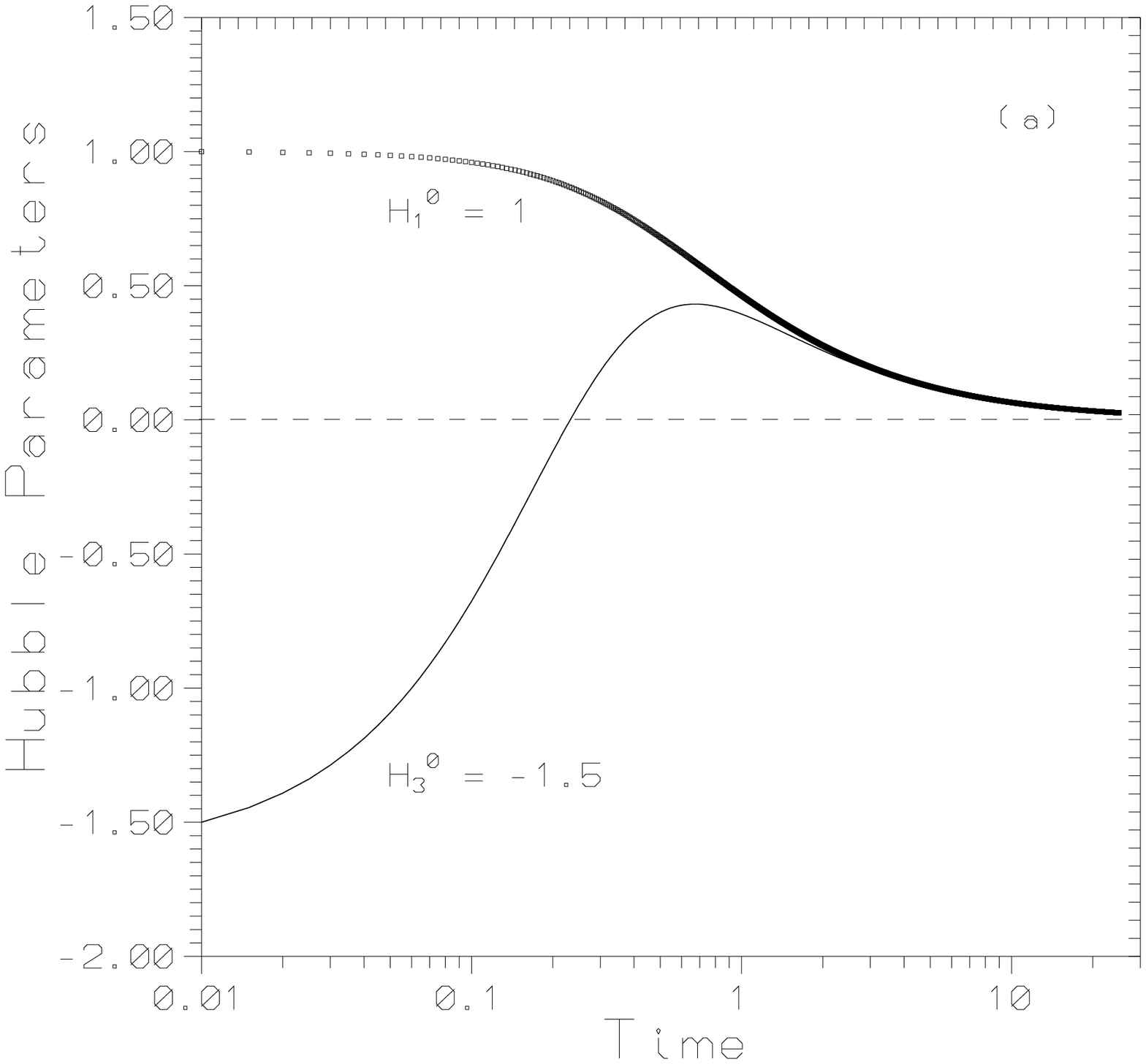}}}
\end{figure}
\begin{figure} 
\centerline{\mbox {\epsfxsize=15.cm \epsfysize=15.cm \epsfbox{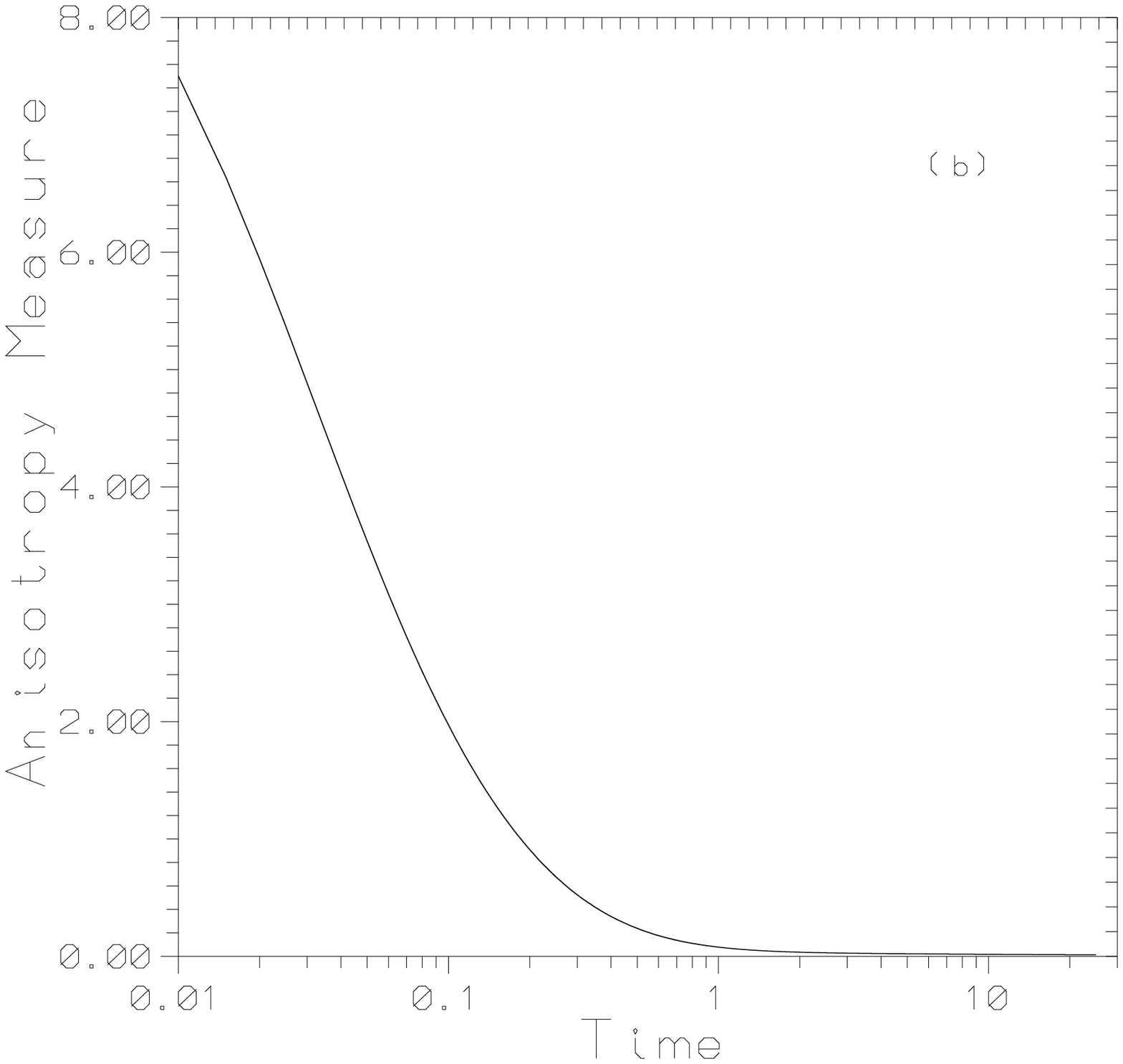}}}
\end{figure}
\begin{figure} 
\centerline{\mbox {\epsfxsize=15.cm \epsfysize=15.cm \epsfbox{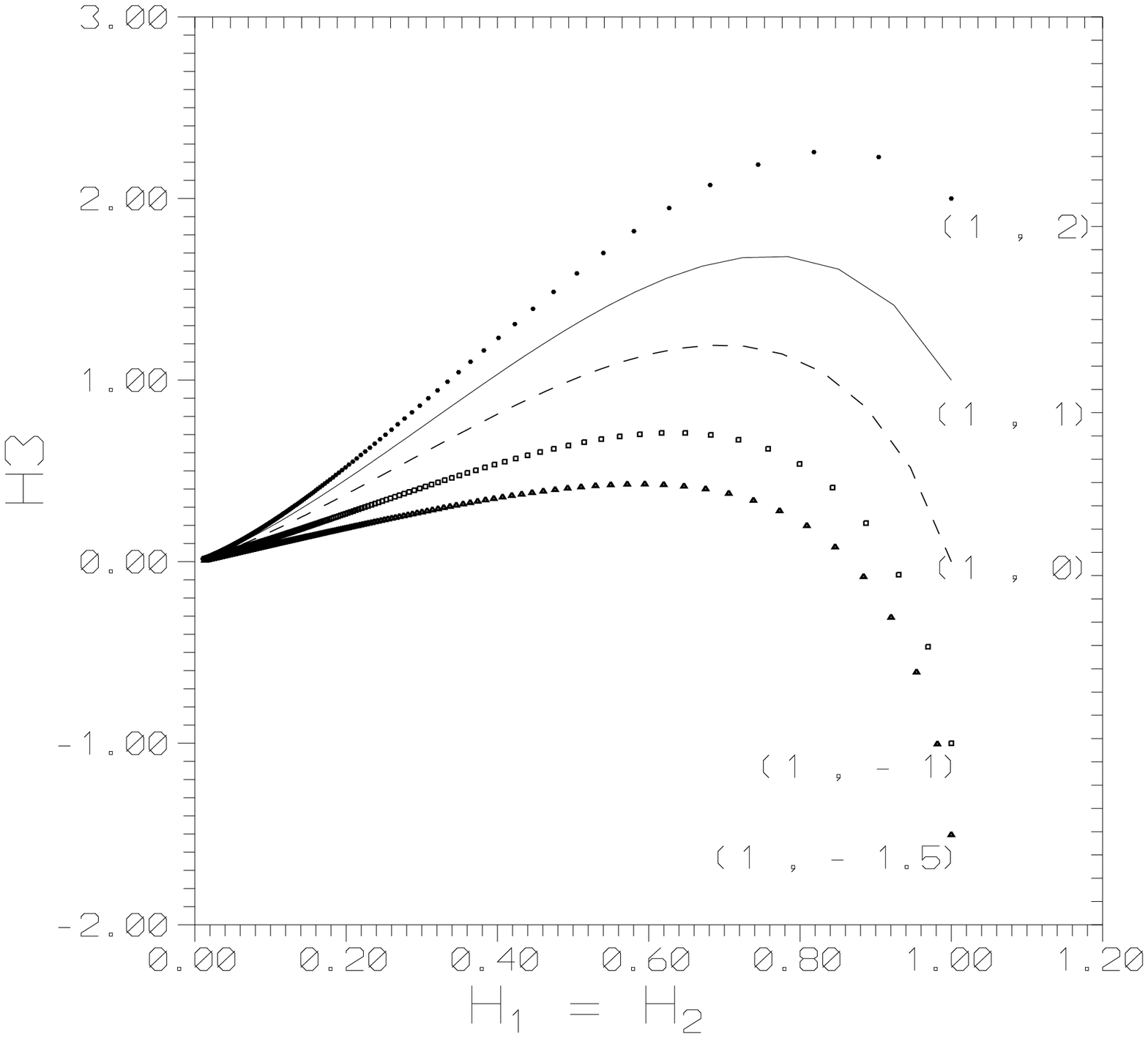}}}
\end{figure}
\begin{figure} 
\centerline{\mbox {\epsfxsize=15.cm \epsfysize=15.cm \epsfbox{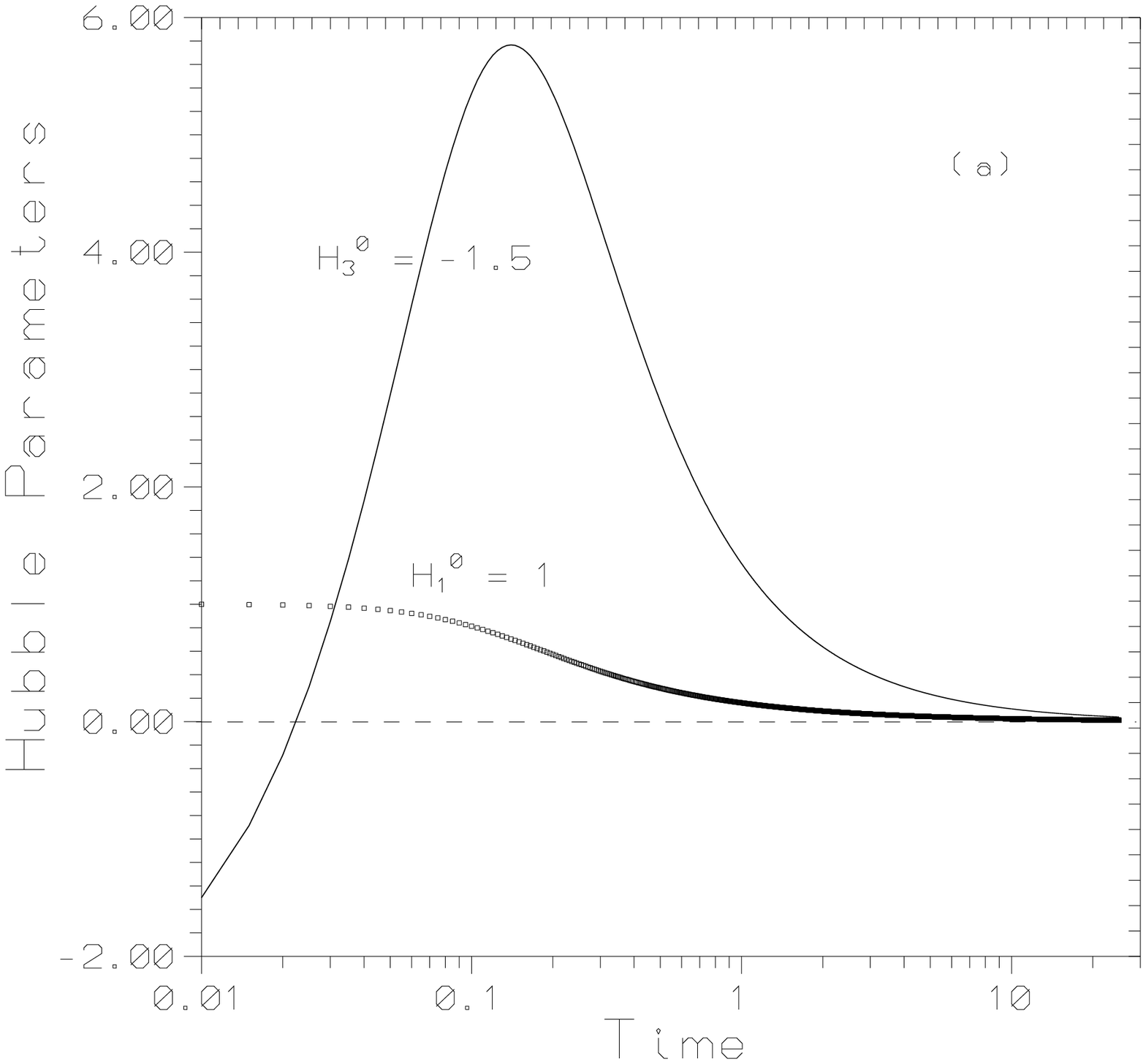}}}
\end{figure}
\begin{figure} 
\centerline{\mbox {\epsfxsize=15.cm \epsfysize=15.cm \epsfbox{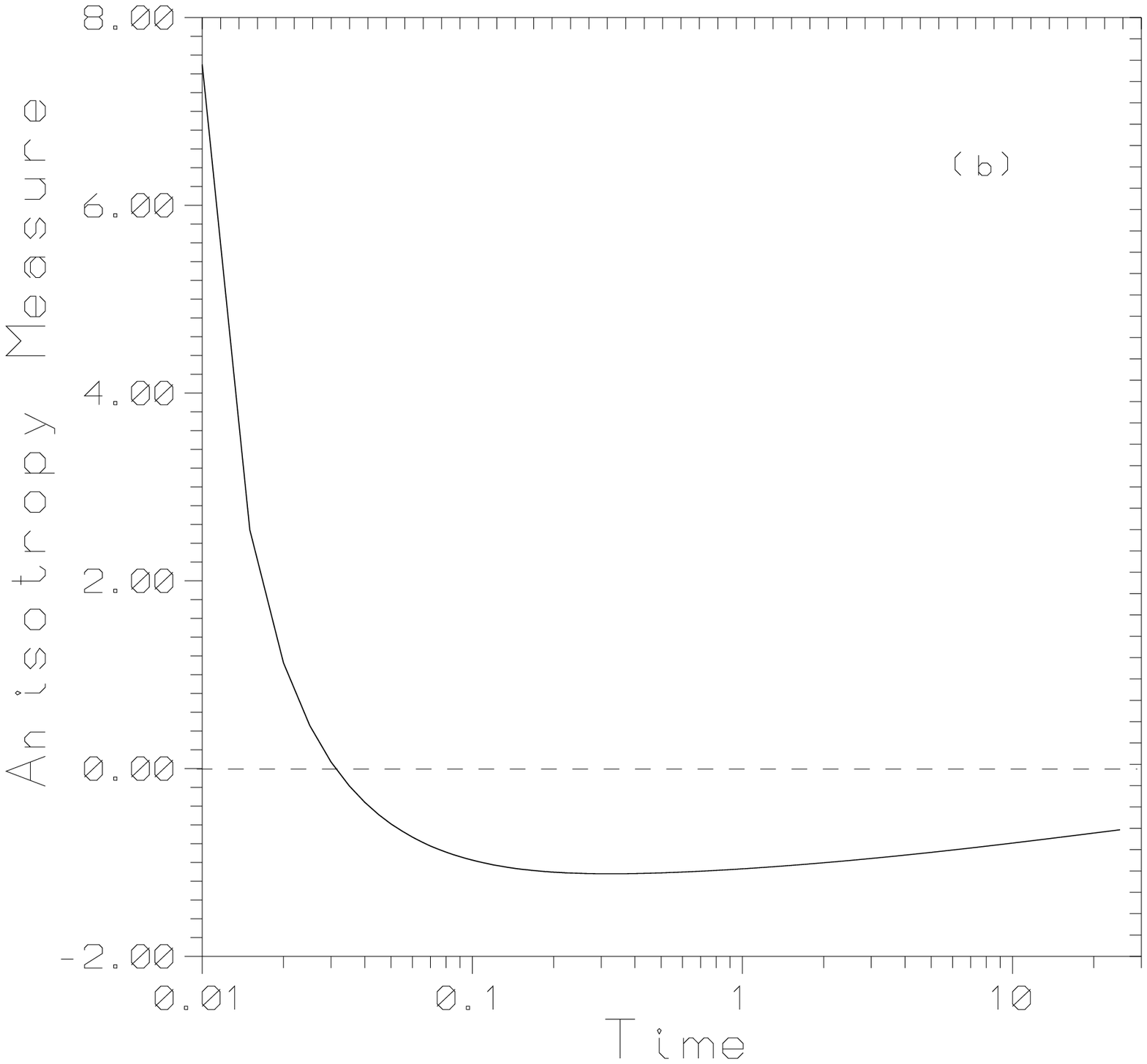}}}
\end{figure}
\begin{figure} 
\centerline{\mbox {\epsfxsize=15.cm \epsfysize=15.cm \epsfbox{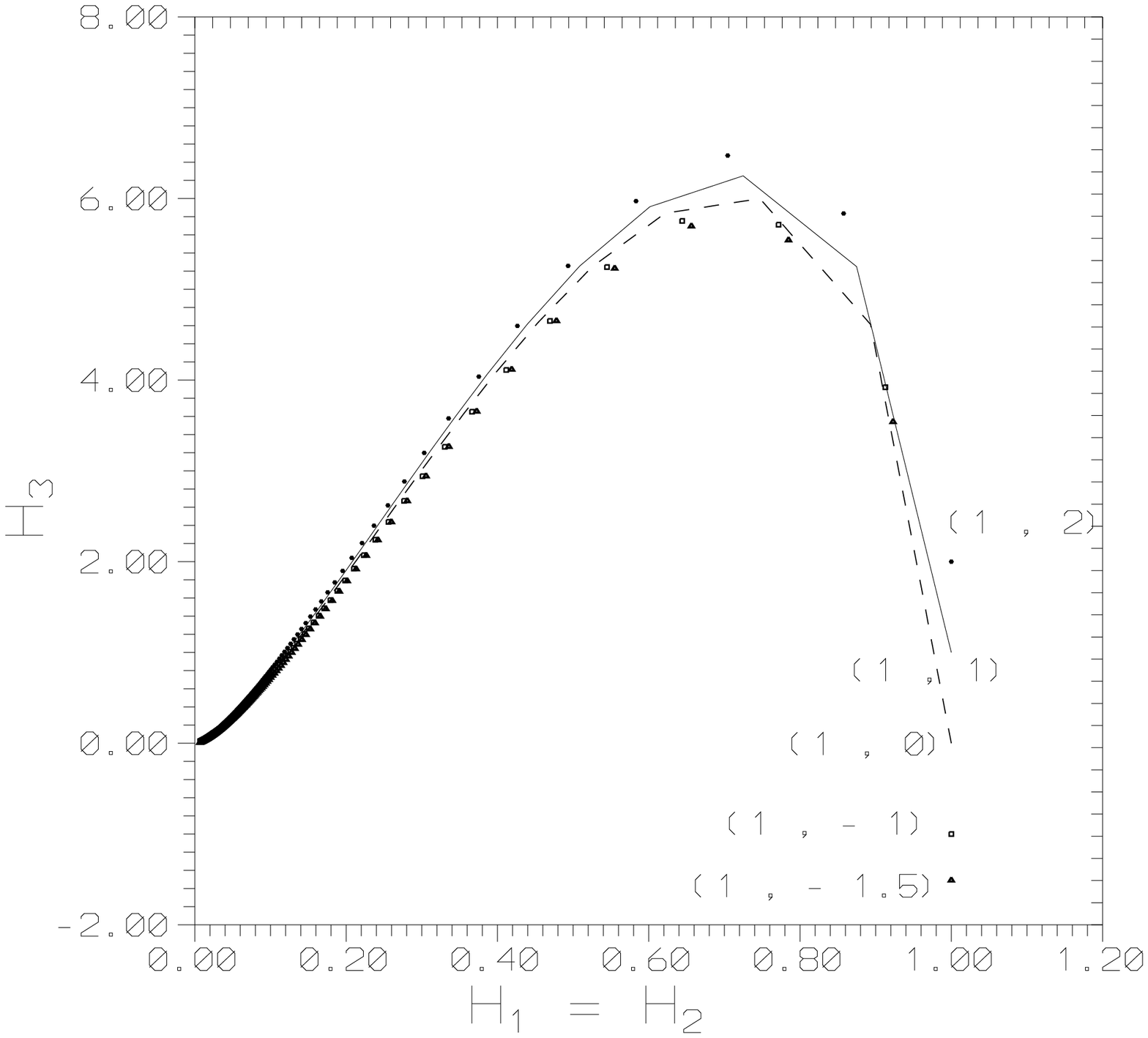}}}
\end{figure}
\begin{figure} 
\centerline{\mbox {\epsfxsize=15.cm \epsfysize=15.cm \epsfbox{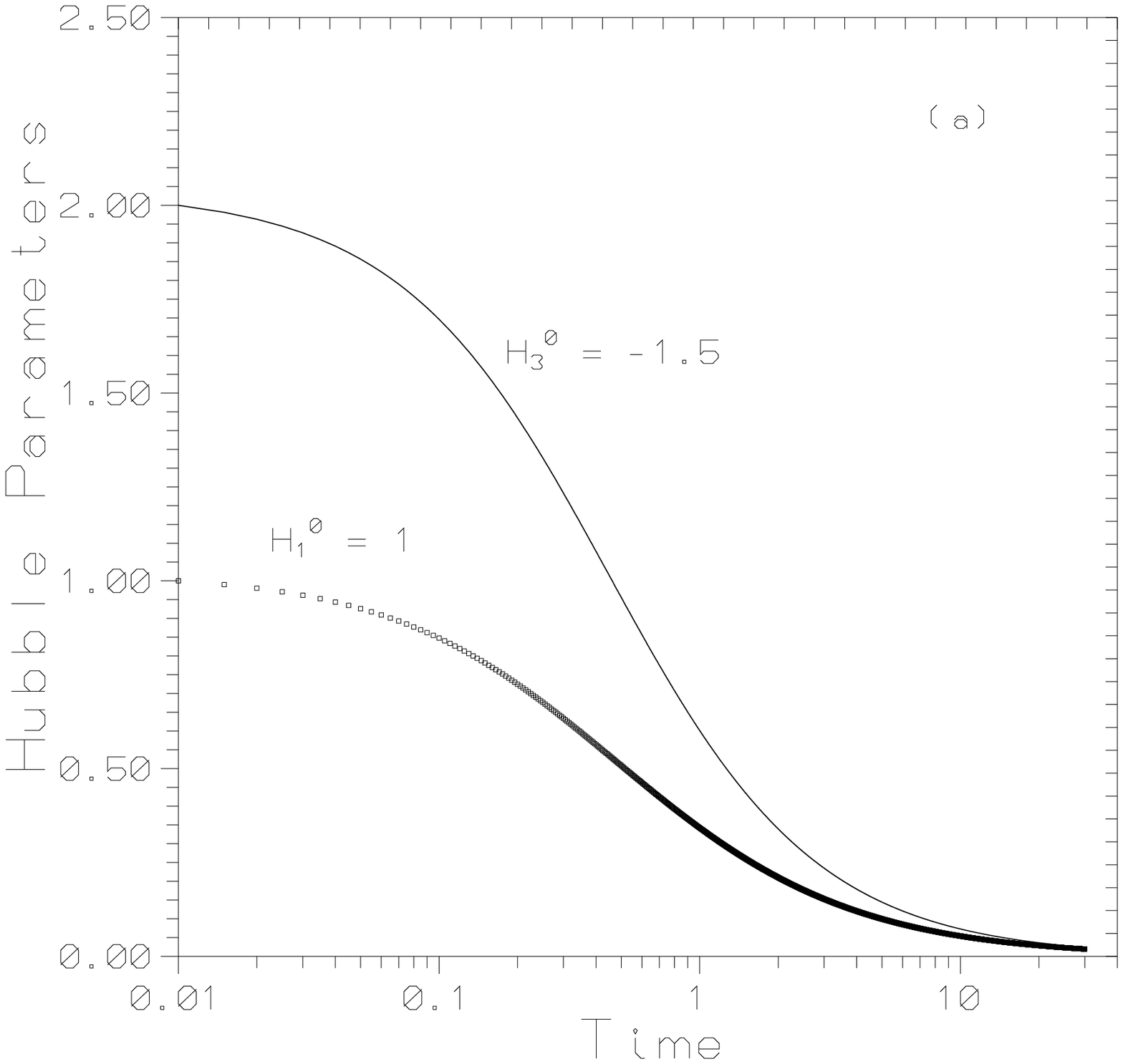}}}
\end{figure}
\begin{figure} 
\centerline{\mbox {\epsfxsize=15.cm \epsfysize=15.cm \epsfbox{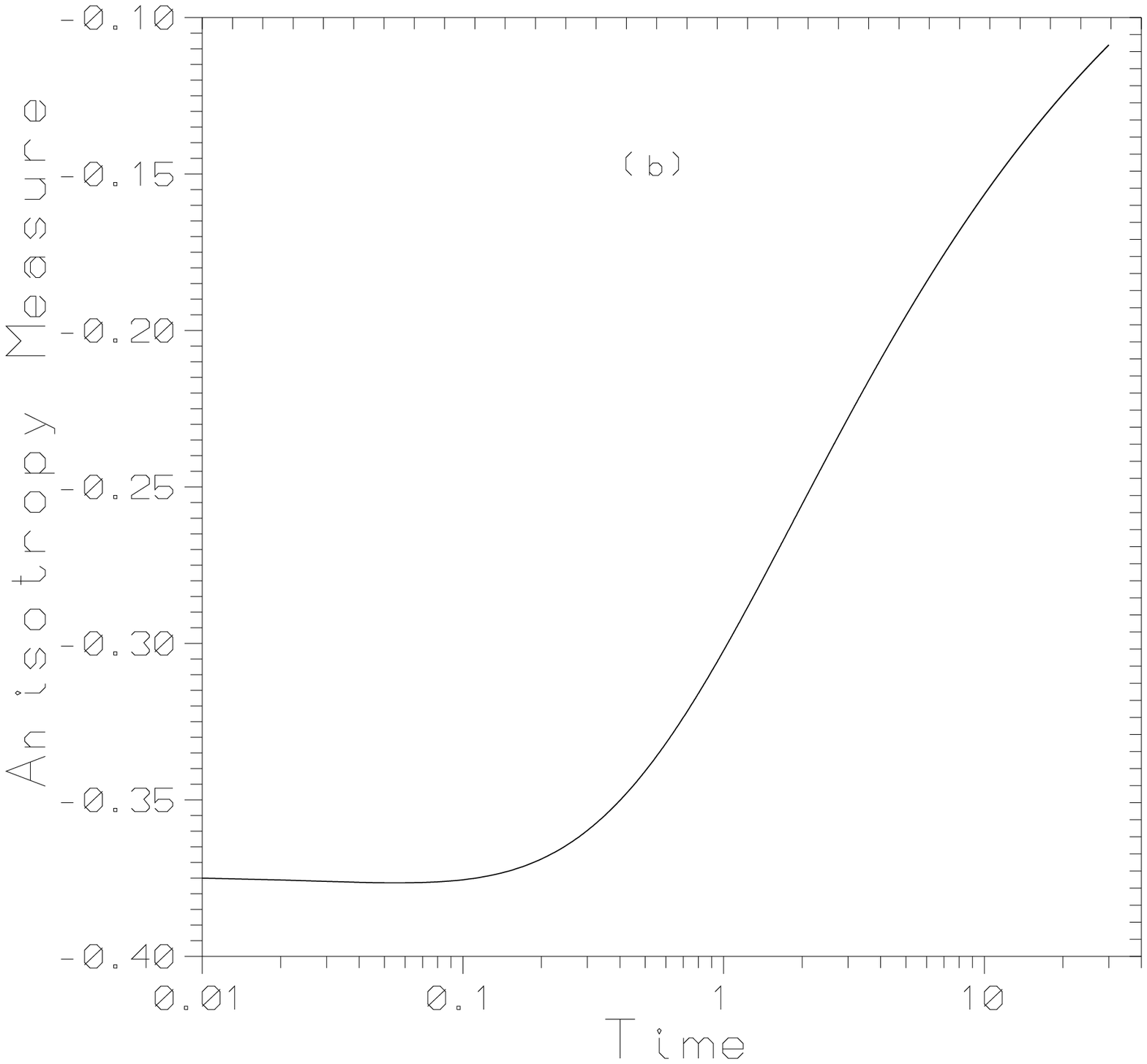}}}
\end{figure}
\begin{figure} 
\centerline{\mbox {\epsfxsize=15.cm \epsfysize=15.cm \epsfbox{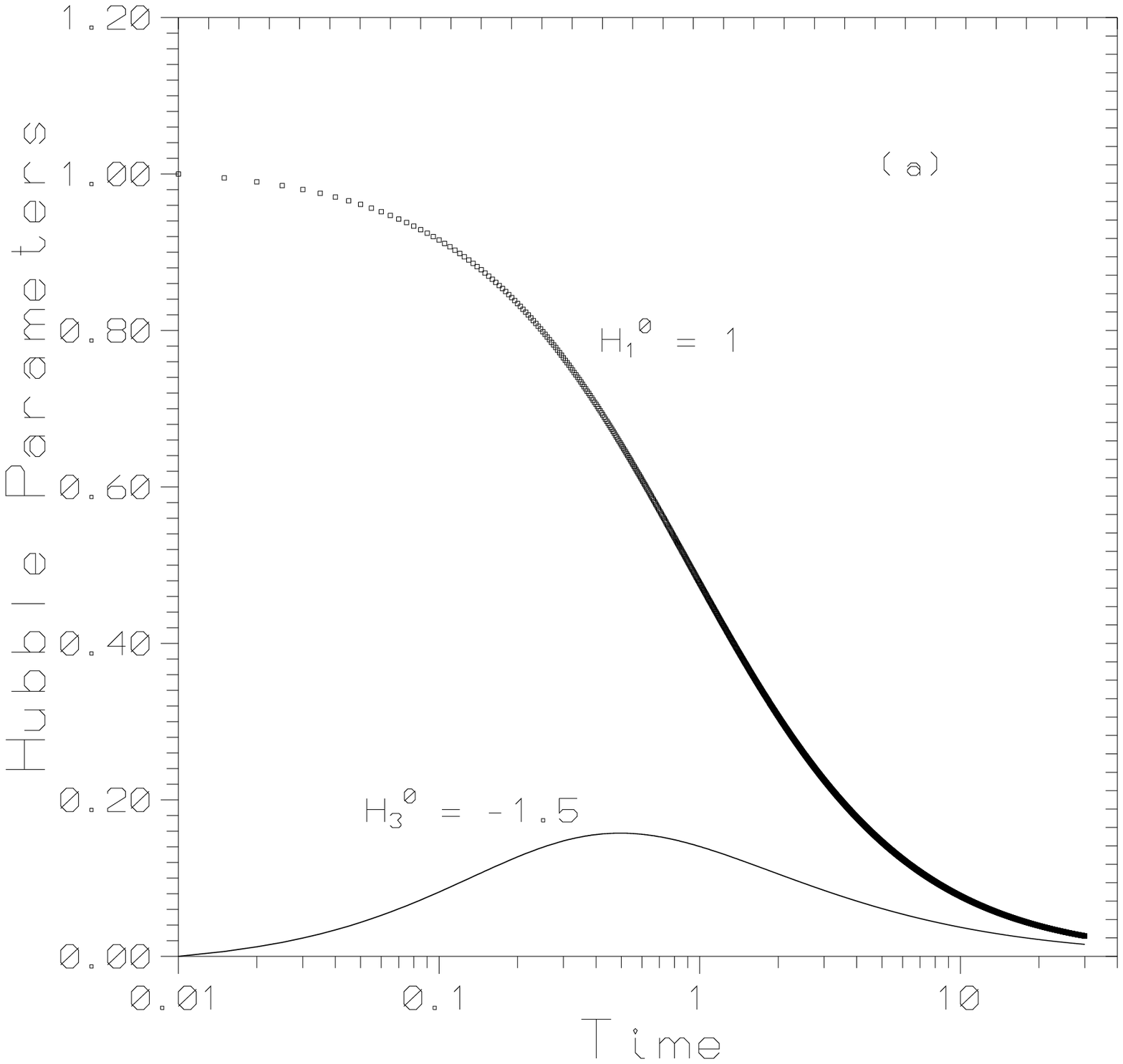}}}
\end{figure}
\begin{figure} 
\centerline{\mbox {\epsfxsize=15.cm \epsfysize=15.cm \epsfbox{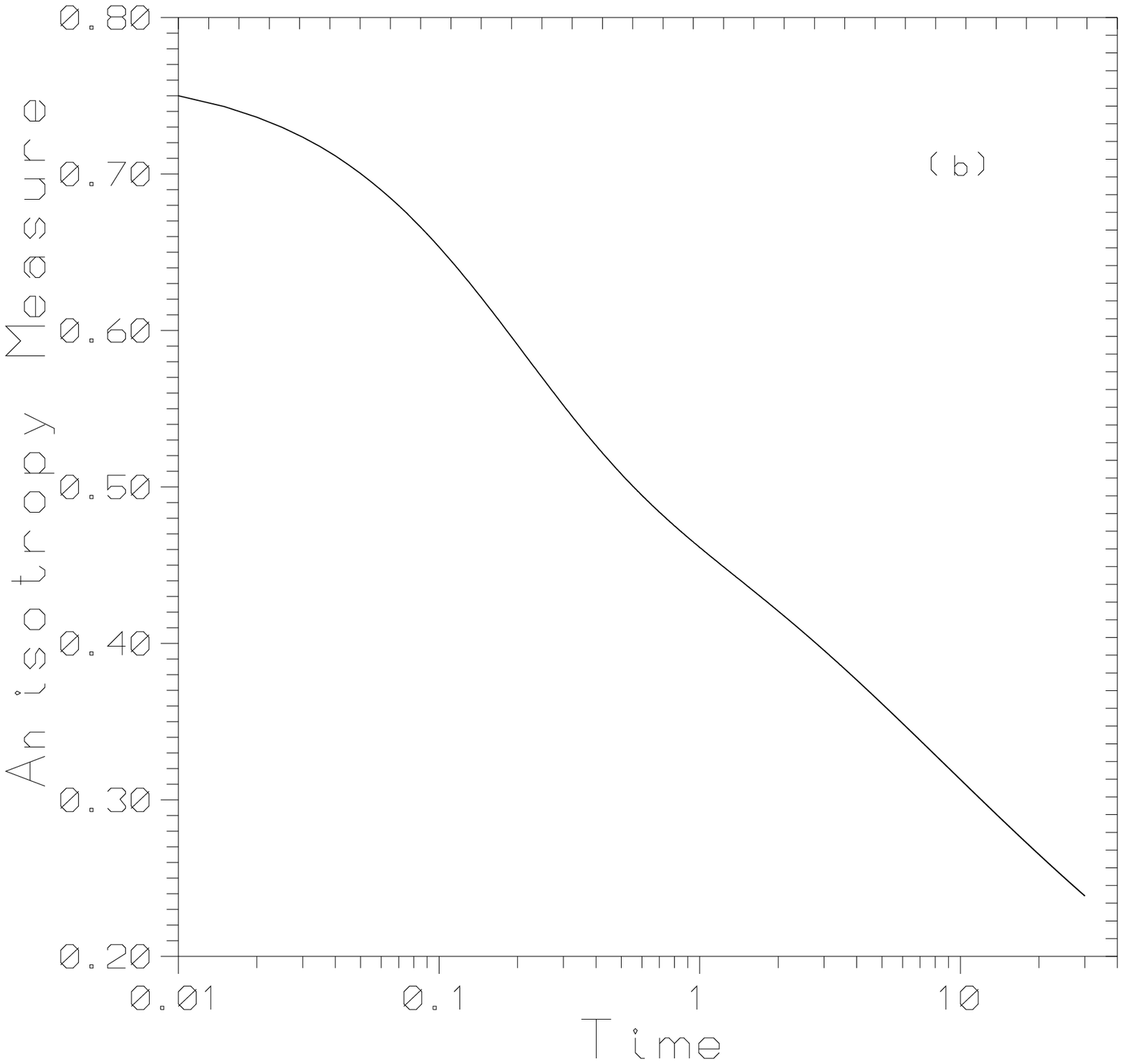}}}
\end{figure}

\end{document}